\documentclass[reprint,twocolumn,secnumarabic,amssymb, aps, prl]{revtex4-1}

\usepackage{graphicx}
\usepackage{amsmath}
\usepackage{braket}
\usepackage[colorlinks,linkcolor=black,citecolor=black,urlcolor=black,filecolor=black]{hyperref}
\usepackage{wrapfig}

\DeclareMathOperator{\sinc}{sinc}

\begin{document}
\definecolor{mygray}{gray}{0.6}
\newcommand\numberthis{\addtocounter{equation}{1}\tag{\theequation}}
\date{\today}

\title{Quantum Phases of Matter on a 256-Atom Programmable Quantum Simulator}


\author{
Sepehr~Ebadi$^{1}$, Tout~T.~Wang$^{1}$, Harry~Levine$^{1}$, Alexander~Keesling$^{1}$, Giulia~Semeghini$^{1}$, Ahmed~Omran$^{1,2}$, Dolev~Bluvstein$^{1}$, Rhine~Samajdar$^{1}$, Hannes~Pichler$^{3,4}$, Wen Wei~Ho$^{1,5}$, Soonwon~Choi$^{6}$, Subir~Sachdev$^{1}$, Markus~Greiner$^{1}$, Vladan~Vuleti\'{c}$^{7}$, and Mikhail~D.~Lukin$^{1}$}
\affiliation{$^1$Department of Physics, Harvard University, Cambridge, MA 02138, USA \\ 
$^2$QuEra Computing Inc., Boston, MA 02135, USA  \quad \quad \quad \quad\\
$^3$Institute for Theoretical Physics, University of Innsbruck, Innsbruck A-6020, Austria \\
$^4$Institute for Quantum Optics and Quantum Information, Austrian Academy of Sciences, Innsbruck A-6020 \\
$^5$ Department of Physics, Stanford University, Stanford, CA 94305, USA. \\
$^6$Department of Physics, University of California Berkeley, Berkeley, CA 94720, USA\\ 
$^7$Department of Physics and Research Laboratory of Electronics, Massachusetts Institute of Technology, Cambridge, MA 02139, USA.
}
 
\begin{abstract}
Motivated by far-reaching applications ranging from quantum simulations of complex processes in physics and chemistry to quantum information processing \cite{PreskillNISQ}, a broad effort is currently underway to build large-scale programmable quantum systems. Such systems provide unique insights into strongly correlated quantum matter \cite{Monroe53Qubit, OpticalLattices2017, BlochMBL2016, GreinerAFM2017,AtomArrayNature2017}, while at the same time enabling new methods for computation \cite{GoogleVQE, SchoelkopfMolecular2020,GoogleSupremacy,PhotonSupremacy} and metrology \cite{QuantumMetrologyReview}. 
Here, we demonstrate a programmable quantum simulator based on deterministically prepared two-dimensional arrays of neutral atoms, featuring strong interactions controlled via coherent atomic excitation into Rydberg states \cite{AntoineReview2020}. Using this approach, we realize a quantum spin model with tunable interactions for system sizes ranging from 64 to 256 qubits. We benchmark the system by creating and characterizing high-fidelity antiferromagnetically ordered states, and demonstrate the universal properties of an Ising quantum phase transition in (2+1) dimensions \cite{SachdevQPT}. We then create and study several new quantum phases that arise from the interplay between interactions and coherent laser excitation \cite{Rhine2020}, experimentally map the phase diagram, and investigate the role of quantum fluctuations. Offering a new lens into the study of complex quantum matter, these observations pave the way for investigations of exotic quantum phases, non-equilibrium entanglement dynamics, and hardware-efficient realization of quantum algorithms. 
\end{abstract}

\maketitle

Recent breakthroughs have demonstrated the potential of programmable quantum systems, with system sizes reaching around fifty trapped ions \cite{Monroe53Qubit,MonroeQAOA2020, Blatt20Qubit} or superconducting qubits \cite{GoogleVQE, SchoelkopfMolecular2020,GoogleSupremacy},
for simulations and computation. Correlation measurements with over seventy photons have been used to perform boson sampling \cite{PhotonSupremacy}, while optical lattices with hundreds of atoms are being used to explore Hubbard models \cite{OpticalLattices2017, BlochMBL2016, GreinerAFM2017}. Larger-scale Ising spin systems have been realized using superconducting elements \cite{DWave2018}, but they lack the coherence essential for probing quantum matter.

Neutral atom arrays have recently emerged as a promising platform for realizing programmable quantum systems  \cite{AntoineReview2020, AntoineTunableNature2016, AtomArrayNature2017}. Based on individually trapped and detected cold atoms in optical tweezers with strong interactions between Rydberg states \cite{WhitlockRydbergReview2020}, atom arrays have been utilized to explore quantum dynamics in one- and two-dimensional systems \cite{AtomArrayNature2017,AntoineAdiabatic2017, WaseemAdiabatic2017, AhnDetailedBalance2018, AntoineTopological2019, Keesling2019}, to create high-fidelity \cite{EndresEntanglement2020}
and large-scale \cite{AtomArrayCats2019} entanglement, to perform parallel quantum logic operations \cite{Saffman2019,AtomArrayPRL2019}, and to realize optical atomic clocks \cite{EndresClock2019,KaufmanClock2020}. While large numbers of atoms have been trapped \cite{KaufmanClock2020} and rearranged in two and three dimensions \cite{Antoine2DSorting, WeissSorting2018, Birkl2019, Antoine3D}, coherent manipulation of programmable, strongly interacting systems with more than a hundred individual particles remains an outstanding challenge. Here, we realize a programmable quantum simulator using arrays of up to 256 neutral atoms with tunable  interactions, demonstrating several novel quantum phases and quantitatively probing the associated phase transitions.

\begin{figure*}
\includegraphics[width=180mm]{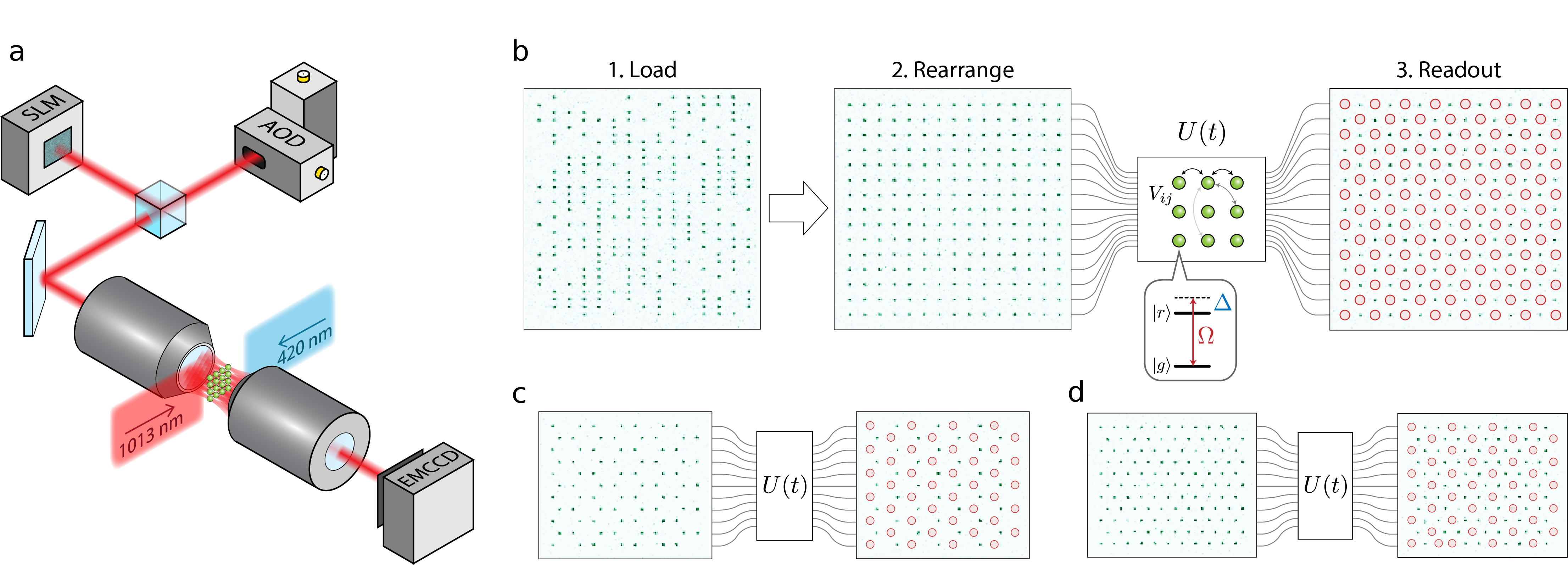}
\caption{\textbf{Programmable two-dimensional arrays of strongly-interacting Rydberg atoms.} \textbf{a.} Atoms are loaded into a 2D array of optical tweezer traps and rearranged into defect-free patterns by a second set of moving tweezers. Lasers at 420~nm and 1013~nm drive a coherent two-photon transition in each atom between ground state $|g\rangle = |5S_{1/2}, F=2, m_F=-2\rangle$ and Rydberg state $|r\rangle = |70S_{1/2}, m_j=-1/2, m_I = -3/2\rangle$. \textbf{b.} Fluorescence image of initial random loading of atoms, followed by rearrangement to a defect-free 15$\times$15 (225 atoms) square array. 
After this initialization, the atoms evolve coherently under laser excitation 
with Rabi frequency $\Omega(t)$ and detuning $\Delta(t)$, and long-range interactions $V_{ij}$. Finally, the state of each atom is read out, with atoms excited to $|r\rangle$ detected as loss and marked with red circles. Shown on the far right is an example measurement following quasi-adiabatic evolution into the checkerboard phase.  \textbf{c, d.} Similar evolution on honeycomb and triangular lattices result in analogous ordered phases of Rydberg excitations with filling $1/2$ and $1/3$, respectively.
}
\label{fig_trapping}
\end{figure*}

\section{Programmable Rydberg Arrays in 2D}
 
Our experiments are carried out on the second generation of an experimental platform described previously \cite{AtomArrayNature2017}. The new apparatus uses a spatial light modulator (SLM) to form a large, two-dimensional (2D) array of optical tweezers in a vacuum cell (Fig.~\ref{fig_trapping}a, Methods). This static tweezer array is loaded with individual $^{87}$Rb atoms from a magneto-optical trap (MOT), with a uniform loading probability of 50--60\% across  up to 1000 tweezers. We rearrange the initially loaded atoms into programmable, defect-free patterns using a second set of moving optical tweezers that are steered by a pair of crossed acousto-optical deflectors (AODs) to arbitrary positions in 2D (Fig.~\ref{fig_trapping}a) \cite{AntoineAssembly2016}. Our parallel rearrangement protocol (see Methods) enables rearrangement into a wide variety of geometries including square, honeycomb, and triangular lattices (left panels in Fig.~1b-d). The procedure takes a total time of $50$--$100$~ms for arrays of up to a few hundred atoms and results in filling fractions exceeding $99\%$.

Qubits are encoded in the electronic ground state 
$|g\rangle$  and the highly-excited $n = 70$ Rydberg state $|r\rangle$ of each atom. 
We illuminate the entire array from opposite sides with two counter-propagating laser beams at 420 and 1013 nm, shaped into light sheets (see Methods), to coherently couple $|g\rangle$ to $|r\rangle$ via a two-photon transition (Fig.~\ref{fig_trapping}a). 

The resulting many-body dynamics $U(t)$ are governed by a combination of the laser excitation and long-range van der Waals interactions between Rydberg states ($V_{ij} = V_0 / |\mathbf{x_i} - \mathbf{x_j}|^6$), described by the Hamiltonian
\begin{equation}
\frac{H}{\hbar} = \frac{\Omega}{2}\sum_i (|g_i\rangle\langle r_i| + |r_i\rangle\langle g_i|) - \Delta\sum_i n_i + \sum_{i<j} V_{ij}n_i n_j
\label{ryd_ham}
\end{equation}
where $\hbar$ is the  Planck's constant, $n_i = |r_i\rangle \langle r_i|$, and $\Omega$ and $\Delta$ are the two-photon Rabi frequency and detuning, respectively. 
After evolution under the Hamiltonian (\ref{ryd_ham}), the state of each atomic qubit is read out by fluorescence imaging that detects only atoms in $|g\rangle$, while atoms in $|r\rangle$
are detected as loss. Detection fidelities exceed 99\% for both states (see Methods).

The Rydberg blockade mechanism \cite{jaksch2000fast,lukin2001} is central to understanding the programmable dynamics driven by the Hamiltonian (\ref{ryd_ham}). It originates from the long-range interactions between Rydberg states, providing an effective constraint that prevents simultaneous excitation of atoms within a blockade radius $R_b\equiv(V_0 / \Omega)^{1/6}$. We  control the effective blockade range $R_b/a$ by programming the lattice spacing $a$ for the atom array. Using these control tools, we explore quantum evolution resulting in a wide variety of quantum phases.

\section{Checkerboard Phase}

\begin{figure}  
\includegraphics[width=\columnwidth]{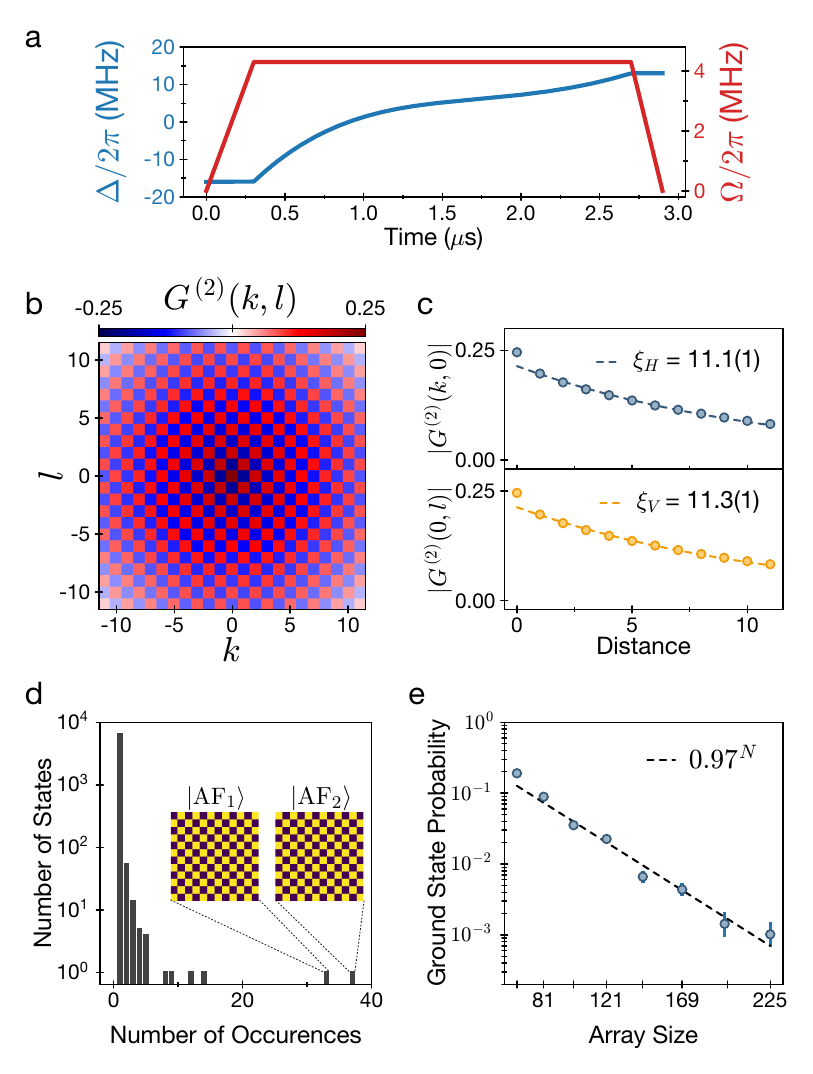}
\caption{\textbf{Benchmarking of quantum simulator using checkerboard ordering.} \textbf{a.} A quasi-adiabatic detuning sweep $\Delta(t)$ at constant Rabi frequency $\Omega$ is used to prepare the checkerboard ground state with high fidelity. \textbf{b.} Two-site correlation function $G^{(2)}(k,l)$, averaged over all pairs of atoms on a 12$\times$12 array, showing near-perfect alternating correlations throughout the entire system. \textbf{c.} Exponential fits of rectified horizontal and vertical correlations are used to extract correlation lengths in the corresponding directions $\xi_H$ and $\xi_V$. \textbf{d.} Histogram of many-body state occurrence frequency after 6767 repetitions of the experiment on a 12$\times$12 array. The two most frequently occurring microstates correspond to the two perfect checkerboard orderings, and the next four most common ones are those with a single defect in one of the corners of the array. \textbf{e.} Probability of finding a perfect checkerboard ground state as a function of array size. The slightly higher probabilities in odd$\times$odd systems is due to commensurate edges on opposing sides of the array. All data in this figure are conditioned on defect-free rearrangement of the array.
}
\label{fig_checkerboard}
\end{figure}

The smallest value of $R_b/a$ that results in an ordered phase for the quantum many-body ground state of the system corresponds to $R_b/a\approx 1$, where only one out of every pair of nearest-neighbor atoms can be excited to $|r\rangle$. On a square array, this constraint leads to a $\mathbb{Z}_2$-symmetry-broken \textit{checkerboard} phase with an antiferromagnetic (AF) ground state. To realize such a state, we initialize the array at $R_b/a = 1.15$ ($a = 6.7~\mu$m, $\Omega = 2\pi \times 4.3$~MHz) with all atoms in $|g\rangle$. We then dynamically sweep the detuning $\Delta$ from negative to positive values while keeping the Rabi frequency $\Omega$ fixed to bring the system quasi-adiabatically into the checkerboard phase (Fig.~\ref{fig_trapping}b and Fig.~\ref{fig_checkerboard}a). A similar approach can be used to create analogous ordered phases on other lattice geometries (Fig.~\ref{fig_trapping}c, d).

We quantify the strength of antiferromagnetic correlations in the checkerboard phase over many experimental repetitions using the connected density-density correlator $G^{(2)}(k,l) = \frac{1}{N_{(k,l)}} \sum_{i,j} (\langle n_i n_j\rangle - \langle n_i\rangle \langle n_j\rangle)$, where the sum is over all pairs of atoms $(i,j)$ separated by the same relative lattice displacement $\mathbf{x}$\,$=$\,$(k,l)$ sites, normalized by the number of such pairs $N_{(k,l)}$. Our measurement of $G^{(2)}(k,l)$ on a 12$\times$12 system (Fig.~\ref{fig_checkerboard}b) yields horizontal and vertical correlation lengths of $\xi_H = $ 11.1(1) and $\xi_V = $ 11.3(1) respectively (Fig.~\ref{fig_checkerboard}c), showing long-range correlations across the entire 144 atom array. These exceed the values reported previously for two-dimensional systems \cite{AntoineAdiabatic2017, WaseemAdiabatic2017} by nearly an order of magnitude.

Single-site readout also allows us to study individual many-body states of our system (Fig.~\ref{fig_checkerboard}d). Out of 6767 repetitions on a 12x12 array, the two perfectly ordered states $|\text{AF}_1\rangle$ and $|\text{AF}_2\rangle$ are by far the most frequently observed microstates, with near-equal probabilities between the two. We benchmark our state preparation by measuring the probability of observing perfect checkerboard ordering as a function of system size (Fig.~\ref{fig_checkerboard}e). We find empirically that the probability scales with the number of atoms according to an exponential $0.97^N$, offering a benchmark that includes all experimental imperfections such as finite detection fidelity, non-adiabatic state preparation, spontaneous emission, and residual quantum fluctuations in the ordered state (see Methods). Remarkably, even for a system size as large as 15$\times$15 (225 atoms), we still observe the perfect antiferromagnetic ground state with probability 0.10${^{+5}_{-4}}$\% within the exponentially large Hilbert space of dimension $2^{225} \approx 10^{68}$. 

\section{(2+1)D Ising Quantum Phase Transition}
We now describe quantitative studies of the quantum phase transition into the checkerboard phase. Quantum phase transitions fall into universality classes characterized by critical exponents that determine \textit{universal} behavior near the quantum critical point, independent of the microscopic details of the Hamiltonian \cite{SachdevQPT}. The transition into the checkerboard phase is expected to be in the paradigmatic---but never previously observed---quantum Ising universality class in (2+1) dimensions \cite{Rhine2020} (with expected dynamical critical exponent $z=1$ and correlation length critical exponent $\nu = 0.629$).

\begin{figure}
\includegraphics[width=\columnwidth]{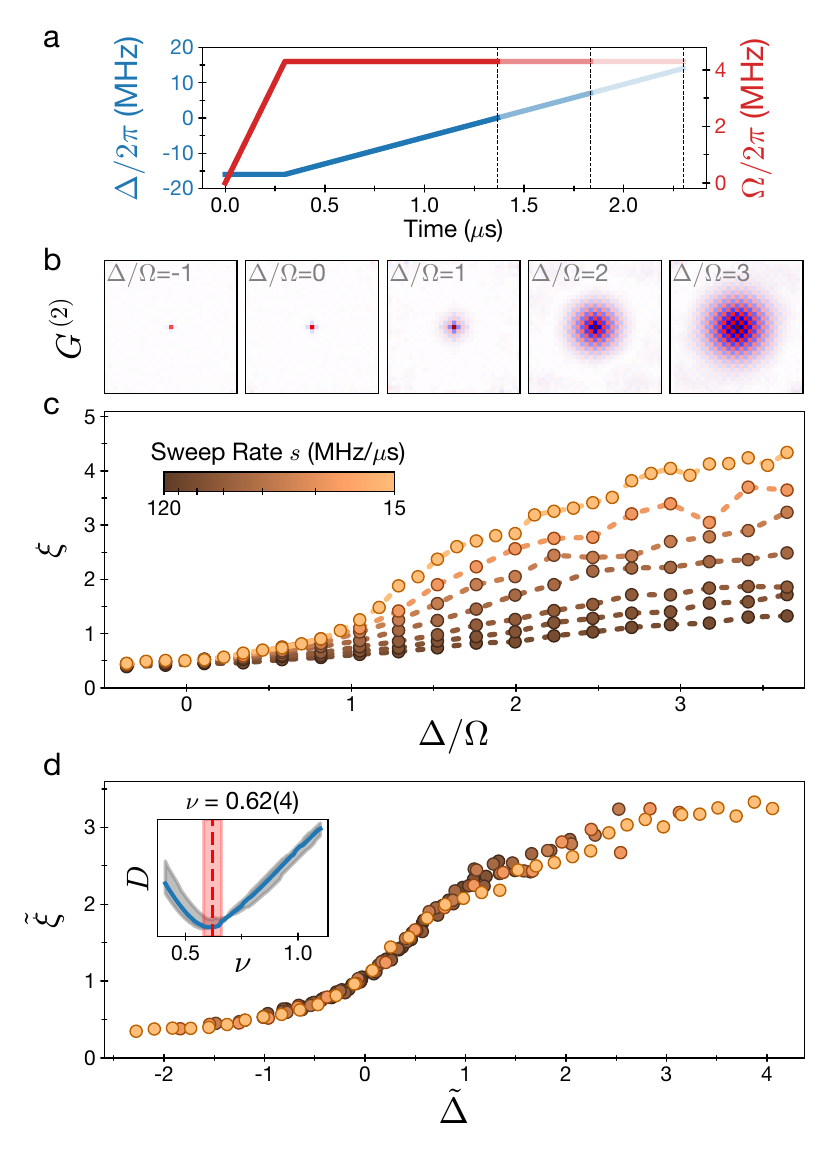}
\caption{\textbf{Observation of the (2+1)D Ising quantum phase transition on a 16$\times$16 array.} 
\textbf{a.} The transition into the checkerboard phase is explored using a linear detuning sweep $\Delta(t)$ at constant $\Omega$. The resulting checkerboard ordering is measured at various endpoints.
\textbf{b.} Example of growing correlations $G^{(2)}$ with increasing $\Delta/\Omega$ along a linear sweep with sweep rate $s = 15$~MHz/$\mu$s. \textbf{c.} Growth of correlation length $\xi$ for $s$ spanning an order of magnitude from 15~MHz/$\mu$s to 120~MHz/$\mu$s. $\xi$ used here measures correlations between the coarse-grained local staggered magnetization (see Methods). \textbf{d.} For an optimized value of the critical exponent $\nu$, all curves collapse onto a single universal curve when rescaled relative to the quantum critical point $\Delta_c$. Inset: distance $D$ between all pairs of rescaled curves as a function of $\nu$ (see Methods). The minimum at $\nu = 0.62(4)$ (red dashed line) yields the experimental value for the critical exponent (red and gray shaded regions indicate uncertainties).
}
\label{fig_kzm}
\end{figure}

To explore universal scaling across this phase transition for a large system, we study the dynamical build-up of correlations associated with the quantum Kibble-Zurek mechanism \cite{QKZM2005, Keesling2019} on a $16\times16$ (256 atoms) array,  at fixed $R_b/a = 1.15$. We start at a large negative detuning with all atoms in $|g\rangle$ and linearly increase $\Delta/\Omega$, stopping at various points to measure the growth of correlations across the phase transition (Fig.~\ref{fig_kzm}a, b). Slower sweep rates $s=d\Delta/dt$ result in longer correlation lengths $\xi$, as expected (Fig.~\ref{fig_kzm}c).
 
The quantum Kibble-Zurek mechanism predicts a universal scaling relationship between the control parameter $\Delta$ and the correlation length $\xi$. Specifically, when both $\Delta$ and $\xi$ are rescaled with the sweep rate $s$ (relative to a reference rate $s_0$)  
\begin{gather}
\tilde{\xi}=\xi(s/s_0)^\mu\\
\tilde{\Delta}=(\Delta-\Delta_c)(s/s_0)^\kappa 
\label{kz_rescale}
\end{gather}
with exponents $\mu \equiv \nu/(1+z\nu)$ and $\kappa \equiv -1/(1+z\nu)$, then universality implies that the rescaled $\tilde{\xi}$ vs. $\tilde{\Delta}$ collapses onto a single curve \cite{Keesling2019} for any sweep rate $s$. Taking $z=1$ to be fixed (as expected for a Lorentz-invariant theory), we extract $\nu$ for our system by finding the value that optimizes this universal collapse.

In order to obtain $\nu$, we first independently determine the position of the critical point $\Delta_c$, which corresponds to the peak of the susceptibility $\chi = - \partial^2 \langle H \rangle /\partial\Delta^2$ and is associated with a vanishing gap \cite{SachdevQPT}.
For adiabiatic evolution under the Hamiltonian (\ref{ryd_ham}), the susceptibility $\chi$ is related to the mean Rydberg excitation density $\langle n \rangle$ by $\chi = \partial \langle  n \rangle / \partial \Delta$ according to the Hellman-Feynman theorem. We measure $\langle n \rangle$ vs. $\Delta$ along a slow linear sweep to remain as adiabatic as possible. We take the numerical derivative of the fitted data to obtain $\chi$, finding its peak to be at $\Delta_c/\Omega = 1.12(4)$  (see Methods). 

Having identified the position of the critical point, we now extract the value of $\nu$ that optimizes data collapse (inset of Fig.~\ref{fig_kzm}d and Methods). The resulting $\nu =0.62(4)$ rescales the experimental data to clearly fall on a single universal curve (Fig.~\ref{fig_kzm}d). This measurement is in good agreement with the predicted $\nu = 0.629$ for the quantum Ising universality class in (2+1) dimensions\cite{Rhine2020}, and distinct from both the mean-field value\cite{SachdevQPT} of $\nu = 1/2$ and the previously verified value in (1+1) dimensions \cite{Keesling2019} of $\nu = 1$. Despite imperfections associated with non-adiabatic state preparation and decoherence in our system, this demonstration of universal scaling highlights opportunities for quantitative studies of quantum critical phenomena on our platform. 

\begin{figure*}[!t]
\includegraphics[width= \textwidth]{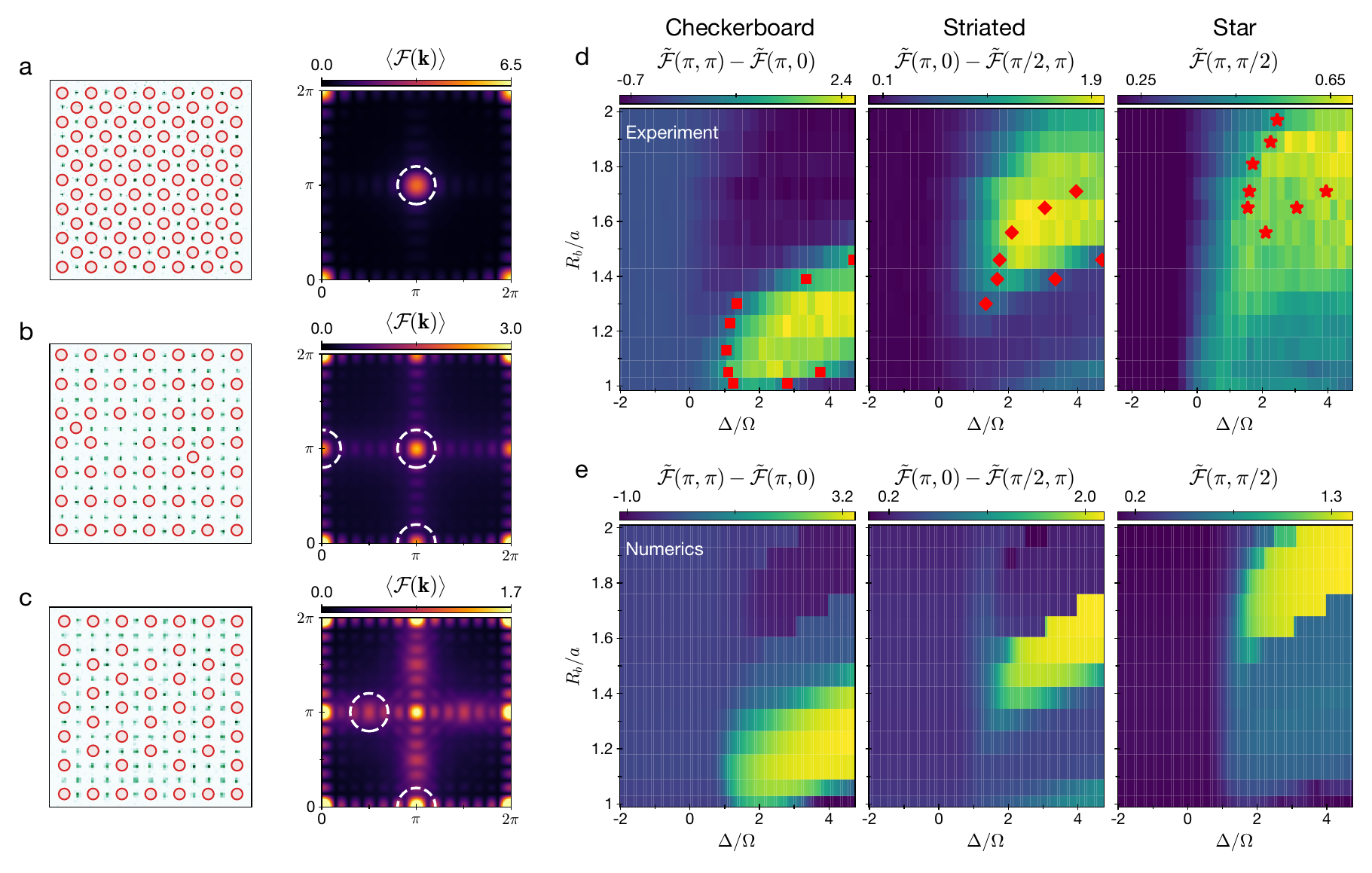}
\caption{\textbf{Phase diagram of the two-dimensional square lattice.} \textbf{a.} Example fluorescence image of atoms in the checkerboard phase and the corresponding Fourier transform averaged over many experimental repetitions $\langle\mathcal{F}(\mathbf{k})\rangle$, highlighting the peak at $(\pi,\pi)$ (circled). \textbf{b.} Image of atoms in the striated phase and the corresponding $\langle\mathcal{F}(\mathbf{k})\rangle$ highlighting peaks at $(0,\pi)$, $(\pi,0)$ and $(\pi,\pi)$ (circled). \textbf{c.} Image of atoms in the star phase with corresponding Fourier peaks at $(\pi/2,\pi)$ and $(\pi,0)$ (circled), as well as at symmetric partners $(\pi,\pi/2)$ and $(\pi,0)$. \textbf{d.} The experimental phase diagram is constructed by measuring order parameters for each of the three phases for different values of the tunable blockade range $R_b/a$ and detuning $\Delta/\Omega$. Red markers indicate the numerically calculated phase boundaries (see Methods). \textbf{e.} The order parameters evaluated numerically using DMRG for a 9$\times$9 array (see Methods).
}
\label{fig_phases}
\end{figure*}

\section{Phase Diagram of the Square Lattice}

A rich variety of new phases have been recently predicted for the square lattice when Rydberg blockade is extended beyond nearest neighbors \cite{Rhine2020}. To map this phase diagram experimentally, we use the Fourier transform of single-shot measurement outcomes $\mathcal{F}(\mathbf{k}) =  \left|\sum_{i} \text{exp} (i \mathbf{k \cdot x_i}/a) n_i / \sqrt{N}\right|$, which characterizes long-range order in our system. For instance, the checkerboard phase shows a prominent peak at $\mathbf{k}=(\pi,\pi)$, corresponding to the canonical antiferromagnetic order parameter: the staggered magnetization (Fig.~\ref{fig_phases}a). We construct order parameters for all observed phases using the symmetrized Fourier transform $ \tilde{\mathcal{F}} (k_1,k_2) = \langle \mathcal{F}(k_1,k_2) + \mathcal{F}(k_2,k_1) \rangle/2$, averaged over experimental repetitions, which takes into account the reflection symmetry in our system (see Methods).

When interaction strengths are increased such that next-nearest (diagonal) neighbor excitations are suppressed by Rydberg interactions ($R_b/a \gtrsim \sqrt{2}$), translational symmetry along the diagonal directions is also broken, leading to the appearance of a new \textit{striated} phase (Fig.~\ref{fig_phases}b). In this phase, Rydberg excitations are mostly located two sites apart and hence appear both on alternating rows and alternating columns. This ordering is immediately apparent through the observation of prominent peaks at $\mathbf{k}~=~(0,\pi)$, $(\pi, 0)$, and $(\pi, \pi)$ in the Fourier domain. As discussed and demonstrated below, quantum fluctuations, appearing as defects on single shot images (Fig.~\ref{fig_phases}b), play a key role in stabilizing this phase. 

At even larger values of $R_b/a \gtrsim 1.7$, the \textit{star} phase emerges, with Rydberg excitations placed every four sites along one direction and every two sites in the perpendicular direction. There are two possible orientations for the ordering of this phase, so Fourier peaks are observed at $\mathbf{k}$ = $(\pi, 0)$ and $(\pi/2, \pi)$, as well as at their symmetric partners $(0, \pi)$ and $(\pi, \pi/2)$ (Fig.~\ref{fig_phases}c). In the thermodynamic limit, the star ordering corresponds to the lowest-energy classical configuration of Rydberg excitations on a square array with a density of 1/4.

We now systematically explore the phase diagram on 13$\times$13 (169 atoms) arrays, with dimensions chosen to be simultaneously commensurate with checkerboard, striated, and star orderings (see Methods). For each value of the blockade range $R_b/a$, we linearly sweep $\Delta$ (similar to Fig.~\ref{fig_kzm}a but with a ramp-down time of 200~ns), stopping at evenly spaced endpoints to raster the full phase diagram.
For every endpoint, we extract the order parameter corresponding to each many-body phase, and plot them separately to show their prominence in different regions of the phase diagram (Fig.~\ref{fig_phases}d).

We compare our observations with numerical simulations of the phase diagram using the density-matrix renormalization group (DMRG) on a smaller 9$\times$9 array with open boundary conditions (Fig.~\ref{fig_phases}e and red markers in Fig.~\ref{fig_phases}d). We find excellent agreement in the extent of the checkerboard phase. For the striated and star phases, we also find good similarity between experiment and theory, although due to their larger unit cells and  the existence of many degenerate configurations, these two phases are more sensitive to both edge effects and experimental imperfections. We emphasize that the numerical simulations evaluate the order parameter for the exact ground state of the system at each point, while the experiment quasi-adiabatically prepares these states via a dynamical process. These results establish the potential of programmable quantum simulators with tunable, long-range interactions for studying large quantum many-body systems that are challenging to access with  state-of-the-art computational tools \cite{Montangero2020}. 

\section{Quantum Fluctuations in the Striated Phase}

\begin{figure}
\includegraphics[width=\columnwidth]{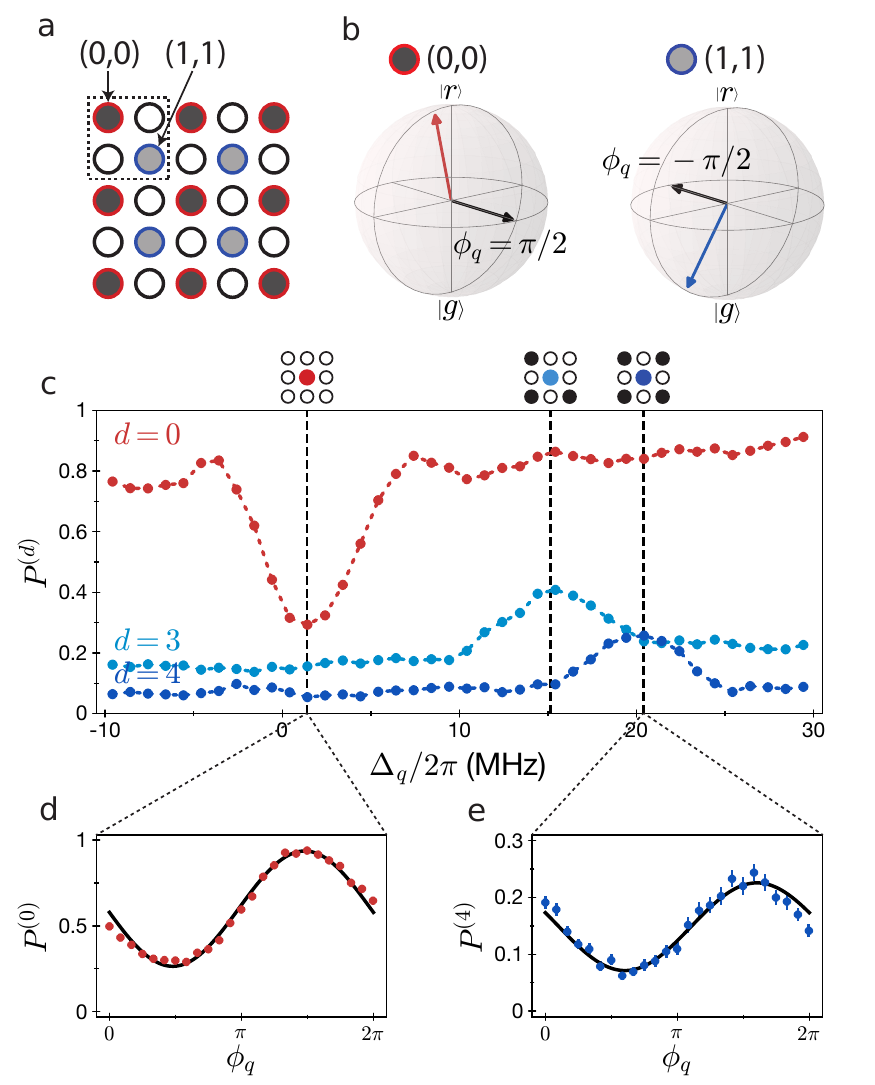}
\caption{\textbf{Probing correlations and coherence in the striated phase via quench dynamics.} \textbf{a.} Unit cell of striated ordering (dashed box) with (0,0) and (1,1) sublattices outlined in red and blue, respectively. The fill shade on each site reflects the mean Rydberg excitation. \textbf{b.} The variational states for the (0,0) and (1,1) sublattices are illustrated  on the Bloch sphere (see Methods). The black arrow illustrates the phase $\phi_q$ of $\Omega$ during the quench. \textbf{c.} Probability $P^{(d)}$ of an excitation, conditioned on observing no nearest-neighbor excitations, and zero (red), three (light blue), or four (dark blue) diagonal next-nearest neighbor excitations. $P^{(0)}$ is plotted for $\phi_q = \pi/2$, showing resonant de-excitation of the (0,0) sublattice near the bare-atom resonance (leftmost vertical line). $P^{(3)}$ and $P^{(4)}$ are plotted for $\phi_q = -\pi/2$, showing excitation peaks for the (1,1) sublattice at interaction shifts corresponding to 3 or 4 diagonal neighbors (two rightmost vertical lines). \textbf{d, e.}  $P^{(0)}$ and $P^{(4)}$ vary with quench phase $\phi_q$ at their corresponding resonances ($\Delta_q/2\pi$ = 1.4 and 20.4~MHz, respectively), demonstrating coherence on both the (0,0) and (1,1) sublattices. Solid line fits are used to extract Bloch vector components.
}
\label{fig_striated}
\end{figure}

We now explore the nature of the striated phase. In contrast to the checkerboard and star phases, which can be understood from a dense-packing argument \cite{Rhine2020}, this phase has no counterpart in the classical limit ($\Omega \to 0$) (see Methods). Striated ordering allows the atoms to lower their energy by partially aligning with the transverse field, favoring this phase at finite $\Omega$. This can be seen by considering the $2\times2$ unit cell, within which one site has a large Rydberg excitation probability (designated the (0,0) sublattice) (Fig.~\ref{fig_striated}a). Excitations on its nearest-neighbor (0,1) and (1,0) sublattices are suppressed due to strong Rydberg blockade. The remaining atoms on the (1,1) sublattice have no nearest neighbors in the Rydberg state and experience a much weaker interaction from four next-nearest (diagonal) neighbors on the (0,0) sublattice, thus allowing the (1,1) atoms to lower their energy by forming a coherent superposition between ground and Rydberg states (Fig.~\ref{fig_striated}b).
 
We experimentally study quantum fluctuations in this phase by observing the response of the system to short quenches (with quench times $t_q < 1/\Omega_q$). The dependence on the detuning $\Delta_q$ and laser phase $\phi_q$ of the quench contains information about local correlations and coherence, which allows us to characterize the quantum states on the different sublattices. The quench resonance for each site depends on the state of its nearest and next-nearest neighbors. Due to the large difference between the interaction energies on the (0,0) and (1,1) sublattices, when one sublattice is resonantly driven, the other is effectively frozen.

The nature of the striated phase is revealed using nine-particle operators to measure the state of an atom, conditioned on its local environment. Specifically, we evaluate the conditional Rydberg density $P^{(d)}$, defined as the excitation probability of an atom if all nearest neighbors are in $|g\rangle$, and exactly $d$ next-nearest (diagonal) neighbors are in $|r\rangle$ (see Methods). For $d=0$, we observe a dip in $P^{(0)}$ near the bare atom resonance (Fig.~\ref{fig_striated}c), corresponding to resonant de-excitation of the (0,0) sublattice. Meanwhile, $P^{(3)}$ and $P^{(4)}$ have two separate peaks that correspond to resonant excitation of the (1,1) sublattice with $d=3$ and $d=4$ next-nearest neighbor excitations, respectively (Fig.~\ref{fig_striated}c). Remarkably, we find that the quench response of both the (0,0) and (1,1) sublattices depends on the phase $\phi_q$ of the driving field during the quench (Fig.~\ref{fig_striated}d,e). The measured visibilities, together with a simple mean-field model (see Methods),  enable the estimation of unknown Bloch vector components on the two sublattices, yielding $\langle \sigma_x\rangle = -0.82(6)$, $\langle \sigma_y\rangle = 0.25(2)$ for the (0,0) sublattice, and $\langle \sigma_x \rangle = -0.45(4)$, $\langle \sigma_y\rangle = 0.09(1)$ for the (1,1) sublattice. We emphasize that accurate characterization requires the use of more sophisticated variational wavefunctions (based on e.g. tensor networks) and warrants further  investigation. This approach can also be  extended through techniques such as shadow tomography \cite{PreskillShadowTomography2020}.


\section{Outlook}

These experiments demonstrate that two-dimensional Rydberg atom arrays constitute a powerful platform for programmable quantum simulations with hundreds of qubits. We expect that system size, quantum control fidelity, and degree of programmability can all be increased considerably via technical improvements. In particular, array sizes and rearrangement fidelities, along with atomic state readout, are currently limited by collisions with background gas particles, and can be improved with an upgraded vacuum system \cite{EndresEntanglement2020} and increased photon collection efficiency. Quantum coherence can be enhanced using higher-power Rydberg lasers and by encoding qubits in hyperfine ground states \cite{AtomArrayPRL2019, WhitlockRydbergReview2020}. Tweezers with different atomic \cite{ThompsonYb2019,EndresEntanglement2020,KaufmanClock2020} and molecular \cite{DoyleMoleculeTweezers2019, NiMoleculeTweezers2019} species can provide additional features and  lead to novel applications in both quantum simulations and metrology. 
Finally, rapidly switchable local control beams can be used to perform universal qubit operations in parallel across the system. 

Our experiments realize several new quantum phases and provide unprecedented insights into quantum phase transitions in two-dimensional systems. These studies can be extended along several directions, including the exploration of non-equilibrium entanglement dynamics via rapid quenches 
across quantum phase transitions \cite{Turner2018, DalmonteLatticeGaugeTheories2020, Scars2020}, the investigation of topological quantum states of matter on frustrated lattices \cite{samajdar2020quantum, verresen2020prediction}, the simulation of lattice gauge theories \cite{LatticeGaugeTheoryReview2020, MontangeroLGT2020}, and the study of broader classes of spin models using hyperfine encoding \cite{ZollerRydbergSimulator2010}. Quantum information processing can also be explored with hardware-efficient methods for multi-qubit operations \cite{AtomArrayPRL2019} and protocols for quantum error correction and fault tolerant control \cite{RydbergFaultTolerance2017}.
Finally, our approach is well suited for efficient implementation of novel algorithms for quantum optimization \cite{FarhiQAOA2014,LeoQAOA} and sampling \cite{ WildSampling2020}, enabling experimental tests of their performance with system sizes exceeding several hundred qubits. 



\vfill\null

\section{Methods}
\noindent\textbf{2D Optical Tweezer Array}\\
Our 2D tweezer array is generated by a free-running 810-nm Ti:Sapphire laser (M Squared, 18-W pump). The laser illuminates a phase-control spatial light modulator (Hamamatsu X13138-02), which imprints a computer generated hologram on the wavefront of the laser field. The phase hologram is calculated using the phase-fixed weighted Gerchberg-Saxton (WGS) algorithm \cite{Kim:19} to produce an arbitrary arrangement of tweezer spots after propagating to the focus of a microscope objective (Mitutoyo: 3.5~mm glass thickness corrected, 50$\times$, NA=0.5). Using this method, we can create tweezer arrays with roughly 1000 individual tweezers (Extended Data Fig.~\ref{fig_large_tweezer_arrays}). When calculating the phase hologram, we improve trap homogeneity by pre-compensating for the variation in diffraction efficiency across the tweezer array (roughly given by $\sinc^2 (\frac{\pi}{2}(\theta_\text{trap}/\theta_\text{max}))$ where $\theta$ denotes the deflection angle from zeroth order).

We also use the phase control of our SLM to correct for optical aberrations on tweezers within the experimentally-used field of view at the plane of atoms (Extended Data Fig.~\ref{fig_tweezer_aberrations}). Aberrations reduce the peak intensity of focal spots (characterized by the Strehl ratio), and correspondingly reduce the light shift of our tweezers on the atoms. By measuring these light shifts as we scan several low-order Zernike polynomials, we quantify and correct for various aberrations in our optical system. Using this method, we compensate for 70 milliwaves of aberrations, observe a total increase of $18\%$ in our trap intensity (Extended Data Fig.~\ref{fig_tweezer_aberrations}c), and measure a corresponding reduction in the range of trap frequencies (Extended Data Fig.~\ref{fig_tweezer_aberrations}d). Aberration correction additionally allows us to place tweezers closer together (minimum separation 3~$\mu$m) to reach larger blockade ranges $R_b/a$.

Tweezers in the array have waists $\sim$~900~nm, trap depths of $\sim~2\pi \times 17$~MHz, and radial trap frequencies of $\sim~2\pi \times ~80$~kHz. In each experimental cycle, the tweezers are loaded from a magneto-optical trap (MOT) with uniform loading probabilities of 50--60\% after 50--100~ms loading time.\\

\noindent\textbf{Atom Rearrangement}\\
Atoms are rearranged using an additional set of dynamically moving tweezers, which are overlaid on top of the SLM tweezer array. These movable tweezers are generated by a separate 809-nm laser source (DBR from Photodigm and tapered amplifier from MOGLabs), and are steered with a pair of independently-controlled crossed acousto-optic deflectors (AODs) (AA Opto Electronic DTSX-400). Both AODs are driven by an arbitrary waveform which is generated in real time using our home-built waveform generation software and an arbitrary waveform generator (AWG)  (M4i.6631-x8 by Spectrum Instrumentation). Dynamically changing the RF frequency allows for continuous steering of beam positions, and  multi-frequency waveforms allow for multiple moving tweezers to be created in parallel \cite{Endres2016}.

While many 2D sorting protocols have been described previously \cite{AntoineAssembly2016, AhnSorting2017, MingshengZhan2DSorting, Birkl2019, Antoine2DSorting}, we implement a novel protocol which is designed to leverage parallel movement of multiple atoms simultaneously. More specifically, we create a row of moving traps which scans upwards along the SLM tweezer array to move one atom within each column up in parallel. This is accomplished by scanning a single frequency component on the vertical AOD to move from the bottom to the top of the SLM array, during which individual frequency components are turned on and off within the horizontal AOD to create and remove tweezers at the corresponding columns. This protocol is designed for SLM tweezer arrays in which traps are grouped into columns and rows. While this does constrain the possible geometries, most lattice geometries of interest can still be defined on a subset of points along fixed columns and rows. \\

\noindent\textbf{Rearrangement Algorithm}\\
Here we detail the rearrangement algorithm, which is illustrated in Extended Data Fig.~3. It operates on an underlying rectangular grid of rows and columns, where the SLM traps correspond to vertices of the grid. We pre-program a set of `target traps' that we aim to fill.

\emph{Pre-sorting:} We begin by ensuring that each column contains a sufficient number of atoms to fill the target traps in that column. In each experimental cycle, due to the random loading throughout the array, some columns may contain excess atoms while other columns may lack a sufficient number of atoms. Accordingly, we apply a `pre-sorting' procedure in which we move atoms between columns. To fill a deficient column $j$, we take atoms from whichever side of $j$ has a larger surplus.
We identify which atoms to take by finding the nearest atoms from the surplus side which are in rows for which column $j$ has an empty trap.
We then perform parallel horizontal sorting to move these atoms into the empty traps of $j$ (not all surplus atoms need to be from the same source column). 

If the one-side surplus is insufficient to fill column $j$, then we move as many surplus atoms as possible from this one side and leave $j$ deficient. We then proceed to the next deficient column, and cycle through until all columns have sufficient atoms.
In typical randomly loaded arrays, this process takes a small number of atom moves compared to the total number of moves needed for sorting. This specific algorithm can fail to properly distribute atoms between columns due to lack of available atoms, but these failures are rare and do not limit the experimental capabilities.

\emph{Ejection:} After pre-sorting, we eject excess atoms in parallel by scanning the vertical AOD frequency downward, beginning at a row in which we want to pick up an atom, and ending below the bottom row of the array.
In each downward scan, we eject a single atom from each column containing excess atoms; we repeat this process until all excess atoms are ejected.

\emph{Parallel sorting within columns:} After pre-sorting and ejection, each column has the correct number of atoms to fill all of its target traps by moving atoms up/down within the column. We now proceed to shuffle the $i^{\text{th}}$-highest loaded atoms to the $i^{\text{th}}$-highest target traps.
As the atoms cannot move through each other, in a single vertical scan atoms are moved as close as possible to their target locations, reaching their targets unless they are blocked by another atom.
We repeat upward/downward scans until all atoms reach their target locations. \\

\noindent\textbf{Rearrangement Parameters and Results}\\
When using moving tweezers to pick up and drop off atoms in the SLM traps, the moving tweezers ramp on/off over $15~\mu$s while positioned to overlap with the corresponding SLM trap. The moving tweezers are approximately twice as deep as the static traps, and  move atoms between SLM traps with a speed of 75~$\mu$m/ms. Typical rearrangement protocols take a total of 50-100~ms to implement in practice, depending on the size of the target array and the random initial loading.
Alignment of the AOD traps onto the SLM array is pre-calibrated by measuring both trap arrays on a monitor CMOS camera and tuning the AOD frequencies to match positions with traps from the SLM array.

A single round of rearrangement results in typical filling fractions of $\sim~98.5\%$ across all target traps in the system. This is limited primarily by the finite vacuum-limited lifetime ($\sim$~10~s) and the duration of the rearrangment procedure. To increase filling fractions, we perform a second round of rearrangement (having skipped ejection in the first round to keep excess atoms for the second round). Since the second round of rearrangement only needs to correct for a small number of defects, it requires far fewer moves and can be performed more quickly, resulting in less background loss. With this approach, we achieve filling fractions of $\sim 99.2\%$ over more than 200 sites, with a total experimental cycle time of $400$~ms.\\

\noindent\textbf{Rydberg Laser System}\\
Our Rydberg laser system is an upgraded version of a previous setup \cite{AtomArrayCats2019}. The 420-nm laser is a frequency-doubled Ti:Sapphire laser (M Squared, 15-W pump). We stabilize the laser frequency by locking the fundamental to an upgraded ultra-low expansion (ULE) reference cavity (notched cylinder design from Stable Laser Systems), with finesse $\mathcal{F}=30,000$ at 840~nm. The 1013-nm laser source is an external-cavity diode laser (Toptica DL Pro), which is locked to the same reference cavity ($\mathcal{F} = 50,000$ at 1013~nm). To suppress high-frequency phase noise from this diode laser, we use the transmitted light through the cavity, which is filtered by the narrow cavity transmission spectrum ($30$~kHz linewidth) \cite{AtomArrayPRL2018}. This filtered light is used to injection-lock another laser diode, whose output is subsequently amplified to 10 W by a fiber amplifier (Azur Light Systems).

Using beam shaping optics to homogeneously illuminate the atom array with both Rydberg lasers, we achieve single-photon Rabi frequencies of $(\Omega_\text{420}, \Omega_\text{1013}) = 2\pi \times (160, 50)~$MHz. We operate with an intermediate state detuning $\delta = 2\pi \times 1$~GHz, resulting in two-photon Rabi frequency $\Omega = \Omega_\text{420} \Omega_\text{1013} / 2\delta \sim 2\pi \times 4$~MHz. Small inhomogeneities in the Rydberg beams result in Rabi frequency variations of $\sim 2\%$ RMS and $\sim 6\%$ peak-to-peak across the array.
With these conditions, we estimate an off-resonant scattering rate of $1/(20~\mu$s) for atoms in $\ket{g}$ and $1/(150~\mu$s) for atoms in $\ket{r}$ at peak power.\\

\noindent\textbf{Rydberg Beam Shaping}\\
We illuminate our 2D atom array with counter-propagating Rydberg laser beams from each side. Instead of using elliptical Gaussian beams, we shape both Rydberg excitation beams into one-dimensional top-hats (light sheets) to homogeneously illuminate the plane of atoms (Extended Data Fig.~\ref{fig_tophat}). To ensure homogenous illumination over the entire array, we define our target field profile in the plane of the atoms with both uniform amplitude cross section and flat phase profile. Using a single phase-only SLM in the Fourier plane to control both phase and amplitude in the image plane is inherently limited in efficiency; therefore, in practice, we compromise between optimizing hologram efficiency and beam homogeneity. We generate these holograms using the conjugate gradient minimization algorithm (Extended Data Fig.~\ref{fig_tophat}c)\cite{Bowman:17}. In all experiments in this work, we use 1D top-hat beams with a flat-width of $105~\mu$m and a perpendicular Gaussian width of $25~\mu$m. The conversion efficiencies into the top-hat modes are 30\% for 420~nm and 38\% for 1013~nm.

Since holographic beam shaping relies on the intricate interplay of different high spatial frequency components in the light field, it is extremely sensitive to optical aberrations. 
We correct for all aberrations up to the window of our vacuum chamber by measuring the amplitude and phase of the electric field as it propagates through the optical beampath (Extended Data Fig.~\ref{fig_tophat}a,b) \cite{Zupancic:16}. We do so by picking off a small portion of the Rydberg beam and observing it on a camera with small pixel size and with sensor cover removed for high-fidelity beam characterization (Imaging Source DMM 27UP031-ML). In this way, we reduce the wavefront error in our beam down to $\lambda/100$, and also use the measured field profile as the starting guess in our hologram generation algorithm (Extended Data Fig.~\ref{fig_tophat}~a,b). Furthermore, by imaging the top-hat beams we also correct for remaining inhomogeneities by updating the input of our optimization algorithm (Extended Data Fig. \ref{fig_tophat}e,f). Due to aberrations and imperfections of the vacuum windows, we observe slightly larger intensity variations on the atoms than expected ($\sim 3\%$ RMS, $\sim 10\%$ peak-to-peak). \\

\noindent\textbf{Rydberg Pulses}\\
After initializing our atoms in the ground state $|g\rangle$, the tweezer traps are turned off for a short time ($<$5~$\mu$s) during which we apply a Rydberg pulse. The pulse consists of a time-dependent Rabi frequency $\Omega(t)$, time-dependent detuning $\Delta(t)$, and a relative instantaneous phase $\phi(t)$. This is implemented by controlling the amplitude, frequency, and phase of the 420-nm laser using a tandem AOM system, similar to what is described previously \cite{AtomArrayCats2019}.\\

\noindent\emph{Quasi-Adiabatic Sweeps:} To prepare many-body ground states with high fidelity, we use an optimized quasi-adiabatic pulse shape (Fig.~\ref{fig_checkerboard}a). The coupling $\Omega(t)$ is initially ramped on linearly at large fixed negative detuning, held constant during the detuning sweep $\Delta(t)$, and finally ramped down linearly at large fixed positive detuning. The detuning sweep $\Delta(t)$ consists of a cubic spline interpolation between five points: initial detuning, final detuning, an inflection point where the slope reaches a minimum, and two additional points that define the duration of the slow part of the sweep. The sweep used for finding perfect checkerboard ground state probabilities (Fig.~\ref{fig_checkerboard}e) was obtained by optimizing the parameters of the spline cubic sweep to maximize the correlation length on a 12$\times$12 (144 atoms) array. The sweep used in detection of the star and striated phases was optimized based on maximizing their respective order parameters. In particular, the inflection point was chosen to be near the position of the minimum gap in these sweeps in order to maximize adiabaticity.  \\

\noindent\emph{Linear Sweeps:}  To probe the phase transition into the checkerboard phase (Fig.~\ref{fig_kzm}), we use variable-endpoint linear detuning sweeps in which $\Omega$ is abruptly turned off after reaching the endpoint.
This ensures that projective readout happens immediately after the end of the linear sweep instead of allowing time for further dynamics,
and is essential for keeping the system within the quantum Kibble-Zurek regime. Linear sweeps are done from $\Delta$ = $-16$ to 14 MHz ($\Delta/\Omega$ = -3.7 to 3.3) at sweep rates $s$ = 15, 21, 30, 42, 60, 85, and 120 MHz/$\mu$s. Data for locating the quantum critical point (Extended Data Fig.~\ref{fig_sus}a) is taken from the slowest of these sweeps ($s$ = 15 MHz/$\mu$s) to remain as close as possible to the ground state. For mapping out the 2D phase diagram (Fig. \ref{fig_phases}), we use the same variable-endpoint linear sweeps at fixed sweep rate $s = 12~$MHz / $\mu$s, except that  $\Omega$ is ramped down over $200~$ns after reaching the endpoint.
\\

\noindent\textbf{State Detection}\\
At the end of the Rydberg pulse, we detect the state of atoms by whether or not they are recaptured in our optical tweezers. Atoms in $\ket{g}$ are recaptured and detected with fidelity $99\%$, limited by the finite temperature of the atoms and collisions with background gas particles in the vacuum chamber.

Atoms excited to the Rydberg state are detected as a loss signal due to the repulsive potential of the optical tweezers on $|r \rangle$. However, the finite Rydberg state lifetime\cite{RydbergProperties2009} ($\sim 80~\mu$s for 70S$_{1/2}$) leads to a probability of $\sim 15\%$ for $|r\rangle$ atoms to decay to $|g\rangle$ and be recaptured by the optical tweezers. In our previous work \cite{AtomArrayCats2019}, we increased tweezer trap depths immediately following the Rydberg pulse to enhance the loss signal for atoms in $\ket{r}$. In 2D, this approach is less effective because atoms which drift away from their initial traps can still be recaptured in a large 3D trapping structure created by out-of-plane interference of tweezers.

Following an approach similar to what has been previously demonstrated\cite{Saffman2019}, we increase the Rydberg detection fidelity using a strong microwave (MW) pulse to enhance the loss of atoms in $|r\rangle$ while leaving atoms in $|g\rangle$ unaffected. The MW source (Stanford Research Systems SG384) is frequency-tripled to $6.9$~GHz and amplified to 3 W (Minicircuits, ZVE-3W-183+). The MW pulse, containing both $6.9$~GHz and harmonics, is applied on the atoms using a microwave horn for $100$~ns. When applying a Rydberg $\pi$-pulse immediately followed by the MW pulse, we observe loss probabilities of $98.6(4)\%$. Since this measurement includes both error in the $\pi$-pulse as well as detection errors, we apply a second Rydberg $\pi$-pulse after the MW pulse, which transfers most of the remaining ground state population into the Rydberg state. In this configuration, we observe $99.1(4)\%$ loss probability, which is our best estimate for our Rydberg detection fidelity (Extended Data Fig~\ref{fig_mw_ionization}). We find that the loss signal is enhanced by the presence of both MW fundamental and harmonic frequencies.\\

\noindent\textbf{Coarse-Grained Local Staggered Magnetization}\\
We define the coarse-grained local staggered magnetization for a site $i$ with column and row indices $a$ and $b$, respectively, as:
$$m_i = \frac{(-1)^{a+b}}{N_i} \sum_{\langle j, i \rangle} (n_i - n_j)$$ where $j$ is summed over nearest neighbors of site $i$ and $N_i$ is the number of such nearest neighbors (4 in the bulk, 3 along the edges, or 2 on the corners).
The value of $m_i$ ranges from $-1$ to 1, with the extremal values corresponding to the two possible perfect antiferromagnetic orderings locally on site $i$ and its nearest neighbors (Extended Data Fig.~\ref{fig_local_stagg}a,b).
The two-site correlation function for $m_i$ can then be defined as an average over experiment repetitions $G^{(2)}_m(k,l) = \frac{1}{N_{(k,l)}} \sum_{i,j} (\langle m_i m_j\rangle - \langle m_i\rangle \langle m_j\rangle)$, where the sum is over all pairs of sites $i, j$ separated by a relative lattice distance of $\mathbf{x} = (k, l)$ sites and normalized by the number of such pairs $N_{(k,l)}$ (Extended Data Fig.~\ref{fig_local_stagg}c). We obtain the correlation length $\xi$ by fitting an exponential decay to the radially averaged $G^{(2)}_m(k,l)$ (Extended Data Fig.~\ref{fig_local_stagg}d). The coarse-grained local staggered magnetization $m_i$ is defined such that the corresponding $G^{(2)}_m(k,l)$ is isotropic (Extended Data Fig.~\ref{fig_local_stagg}c), which makes for natural radial averaging. This radial average captures correlations across the entire array better than purely horizontal or vertical correlation lengths $\xi_H$ and $\xi_V$, which are more sensitive to edge effects.\\

\noindent\textbf{Determination of the Quantum Critical Point}\\
To accurately determine the location of the quantum critical point $\Delta_c$ for the transition into the checkerboard phase, we measure mean Rydberg excitation $\langle n \rangle$ vs. detuning $\Delta$ for a slow linear sweep with sweep rate $s = 15$~MHz/$\mu$s (Extended Data Fig. \ref{fig_sus}a). To smoothen the measured curve, we fit a polynomial for $\langle n \rangle$ vs. $\Delta$ and take its numerical derivative to identify the peak of the susceptibility $\chi$ as the critical point\cite{SachdevQPT} (Extended Data Fig.~\ref{fig_sus}b).

Small oscillations in $\langle n \rangle$ result from the linear sweep not being perfectly adiabatic. To minimize the effect of this on our fitting, we use the lowest-degree polynomial (cubic) whose derivative has a peak, and choose a fit window in which the reduced chi-squared metric indicates a good fit. Several fit windows around $\Delta/\Omega = 0$ to 2 give good cubic fits, and we average results from each of these windows to obtain $\Delta_c/\Omega$ = 1.12(4).

We also numerically extract the critical point for a system with numerically-tractable dimensions of 10$\times$10. Using the density-matrix renormalization group (DMRG) algorithm, we evaluate $\langle n \rangle$ as a function of detuning $\Delta$, and then take the derivative to obtain a peak of the susceptibility at $\Delta_c/\Omega = 1.18$ (Extended Data Fig.~\ref{fig_sus}c,d). To corroborate the validity of our experimental fitting procedure, we also fit cubic polynomials to the DMRG data and find that the extracted critical point is close to the exact numerical value (Extended Data Fig.~\ref{fig_sus}d). This numerical estimate of the critical point for a 10$\times$10 array is consistent with the experimental result on a larger $16\times 16$ array. Moreover, our experiments on arrays of different sizes show that $\Delta_c/\Omega$ does not vary significantly between $12\times 12$, $14\times 14$, and $16\times 16$ arrays (Extended Data Fig. \ref{fig_collapse_distance}b).\\

\noindent\textbf{Data Collapse for Universal Scaling}\\
Optimizing the universal collapse of rescaled correlation length $\tilde{\xi}$ vs. rescaled detuning $\tilde{\Delta}$ requires defining a measure of the distance between rescaled curves for different sweep rates $s_i$. Given $\tilde{\xi}^{(i)}_j$ and $\tilde{\Delta}^{(i)}_j$, where the index $i$ corresponds to sweep rate $s_i$ and $j$ labels sequential data points along a given curve, we define a distance \cite{Seno2001}
\begin{equation}
D = \sqrt{\frac{1}{N}\sum_i \sum_{i'\neq i} \sum_j \left \lvert\tilde{\xi}^{(i')}_j - f^{(i)}\left(\tilde{\Delta}^{(i')}_j\right)\right \rvert^2}.
\end{equation}
The function $f^{(i)}(\tilde{\Delta})$ is the linear interpolation of $\tilde{\xi}^{(i)}_j$ vs. $\tilde{\Delta}^{(i)}_j$, while $N$ is the total number of terms in the three nested sums. The sum over $j$ only includes points that fall within the domain of overlap of all data sets, avoiding the problem of linear interpolation beyond the domain of any single data set. Defined in this way, the collapse distance $D$ measures all possible permutations of how far each rescaled correlation growth curve is from curves corresponding to other sweep rates.

Applied to our experimental data, $D$ is a function of both the location of the critical point $\Delta_c$ and the critical exponent $\nu$ (Extended Data Fig. \ref{fig_collapse_distance}a). Using the independently measured $\Delta_c/\Omega = 1.12(4)$, we obtain $\nu = 0.62(4)$ for optimal data collapse, and illustrate in particular the better collapse for this value than for other values of $\nu$ (Extended Data Fig.~\ref{fig_collapse_distance}c-e). The quoted uncertainty is dominated by the corresponding uncertainty of the extracted $\Delta_c/\Omega$, rather than by the precision of finding the minimum of $D$ for a given $\Delta_c/\Omega$. Our experiments give consistent values of $\Delta_c/\Omega$ and $\nu$ for systems of size 12$\times$12, 14$\times$14, and 16$\times$16 (Extended Data Fig.~\ref{fig_collapse_distance}b).\\

\noindent\textbf{Order Parameters for Many-Body Phases}\\
We construct order parameters to identify each phase using the Fourier transform to quantify the amplitude of the observed density-wave ordering. We define the symmetrized Fourier transform $ \tilde{\mathcal{F}} (k_1,k_2) = \langle \mathcal{F}(k_1,k_2) + \mathcal{F}(k_2,k_1) \rangle/2$ to take into account the $\textit{C}_4$ rotation symmetry between possible ground-state orderings for some phases. For the star phase, the Fourier amplitude $\tilde{\mathcal{F}} (\pi, \pi/2)$ is a good order parameter because ordering at $\mathbf{k} = (\pi, \pi/2)$ is unique to this phase. The striated phase, on the other hand, shares its Fourier peaks at $\mathbf{k}$ = $(\pi, 0)$ and $(0, \pi)$ with the star phase, and its peak at $\mathbf{k}$ = $(\pi, \pi)$ with the checkerboard phase; hence, none of these peaks alone can serve as an order parameter. 
We therefore construct an order parameter for the striated phase to be $\tilde{\mathcal{F}} (0, \pi) - \tilde{\mathcal{F}} (\pi/2, \pi)$, which is nonzero in the striated phase and zero in both checkerboard and star. Similarly, the checkerboard shares its $\mathbf{k} = (\pi, \pi)$ peak with the striated phase, so we construct $\tilde{\mathcal{F}} (\pi,\pi) - \tilde{\mathcal{F}} (0, \pi)$ as an order parameter which is zero in the striated phase and nonzero only in checkerboard. \\

\noindent\textbf{Numerical Simulations of the 2D Phase Diagram}\\
We numerically compute the many-body ground states at different points in the $(\Delta/\Omega, R_b/a)$ phase diagram using the density-matrix renormalization group (DMRG) algorithm \cite{white1992density,white1993density}, which operates in the space of the so-called matrix product state (MPS) ans\"{a}tze.
While originally developed for one-dimensional systems, DMRG can also be extended to two dimensions by representing the 2D system as a winding 1D lattice \cite{stoudenmire2012studying}, albeit with long-range interactions. A major limitation to two-dimensional DMRG is that the number of states required to faithfully represent the ground-state wavefunction has to be increased exponentially with the width of the system in order to maintain a constant accuracy. For our calculations, we employ a maximum bond dimension of $1600$, which allows us to accurately simulate $10\times 10$ square arrays \cite{Rhine2020}. We also impose open boundary conditions in both directions and truncate the van der Waals interactions so as to retain up to third-nearest-neighbor couplings.  
The numerical convergence criterion is set by the truncation error, and the system is regarded to be well-converged to its true ground state once this error drops below a threshold of $10^{-7}$. In practice, this was typically found to be achieved after $\mathcal{O}(10^2)$ successive sweeps.

Since the dimensions of the systems studied in Figure~\ref{fig_phases}, (13\,$\times$\,13 (experimentally) and 9\,$\times$\,9 (numerically), are both of the form $(4n+1)$\,$\times$\,$(4n+1)$, the two phase diagrams are expected to be similar. In particular, both these system sizes are compatible with the commensurate ordering patterns of the crystalline phases observed in this work, and can host all three phases (at the appropriate $R_b$/a) with the same boundary conditions. Likewise, for extraction of the QCP, we use a 10$\times$10 array as it is the largest numerically accessible square lattice comparable to the 16$\times$16 array used in our study of the quantum phase transition.\\

\noindent\textbf{Mean-Field Wavefunction for the Striated Phase}\\
To understand the origin of the striated phase, it is instructive to start from a simplified model in which we assume that nearest-neighbor sites are perfectly blockaded. Since we always work in a regime where $R_b/a > 1$, this model should also capture the essential physics of the full Rydberg Hamiltonian.

In the classical limit of $\Omega = 0$, the perfect checkerboard state has an energy per site of $-\Delta/2 + V (\sqrt{2}a) + V (2a)$, with $V(x)$ being the interaction between sites at a distance $x$, whereas the corresponding energy for the star-ordered state is $-\Delta/4$ (neglecting interactions for $x>2a$). Accordingly, there is a phase transition between the checkerboard and star phases when $\Delta = 4 [V (\sqrt{2}a) + V (2a)]$. On the other hand, for the same density of Rydberg excitations, the striated phase has a classical energy per site of $-\Delta/4 + V (2a)/2$, which is always greater than that of the star phase; hence, striated ordering never appears in the classical limit.

At finite $\Omega$, however, the striated phase emerges due to a competition between the third-nearest-neighbor interactions and the second-order energy shift upon dressing a ground state atom off-resonantly with the Rydberg state. We can thus model the ground state of the striated phase as a product state, where (approximately) $1/2$ of the atoms are in the ground state, $1/4$ of the atoms are in the Rydberg state, and the remaining $1/4$ are in the ground state with a weak coherent admixture of the Rydberg state. A general mean-field ansatz for a many-body wavefunction of this form is given by
\begin{alignat}{1}
\rvert \Psi^{}_\textsc{str} (a_1^{}, a^{}_2) \rangle = &\bigotimes_{\mathbf{i} \in A_1} \left ( \cos a_1 \rvert g \rangle_\mathbf{i} + \sin a_1 \rvert r \rangle_\mathbf{i} \right)\\
\nonumber&\bigotimes_{\mathbf{i} \in A_2} \left ( \cos a_2 \rvert g \rangle_\mathbf{i} + \sin a_2 \rvert r \rangle_\mathbf{i} \right) \bigotimes_{\mathbf{j} \in B} \rvert g \rangle_\mathbf{j},
\end{alignat}
where $A_1$ and $A_2$ represent the two sublattices of the (bipartite) $A$ sublattice, and $a_{1,2}$ are variational parameters. If $a_1=a_2$, then our trial wavefunction simply represents a checkerboard state, but if $a_1\ne a_2$, this state is \textit{not} of
the checkerboard type, and leads to the striated phase.

Based on this ansatz, we can now explicitly see how the striated phase may become energetically favorable in the presence of a nonzero $\Omega$. Consider the atoms on the partially excited sublattice to be in the superposition $\rvert g \rangle + [\Omega/\{4V(\sqrt{2}a)-\Delta \}]\rvert r \rangle$; this describes the state of the atoms on the $(1,1)$ sublattice in the notation of Fig.~\ref{fig_striated}. The net energy per site of the system is then
\begin{equation*}
-\frac{\Delta}{4} + \frac{V(2a)}{2} -\frac{\Omega^2}{ 4\,(4V(\sqrt{2}a) -\Delta)} +\frac{\Omega^2\, V(\sqrt{2}a)}{ 2\,(4V(\sqrt{2}a) -\Delta)^2}
\end{equation*}
where the third and fourth terms are the second-order energy shift and mean-field interaction shift, respectively.
From this expression, we observe that if the energy gained from the dressing (these last two terms) is larger than $V(2a)/2$, then the striated phase prevails over the star phase.\\

\noindent\textbf{Dynamical Probe of the Striated Phase}\\
We prepare striated ordering using an optimized cubic spline sweep along $R_b/a = 1.47$, ending at $\Delta/\Omega = 2.35$. Immediately after this sweep, the system is quenched to detuning $\Delta_q$ and relative laser phase $\phi_q$. We quench at a lower Rabi frequency $\Omega_q = \Omega/4 \approx 2\pi \times 1$~MHz to improve the resolution of this interaction spectroscopy. For the chosen lattice spacing, the interaction energy between diagonal excitations is $2\pi \times 5.3~$MHz. The reference phase for the atoms $\phi = 0$ is set by the instantaneous phase of the Rydberg coupling laser at the end of the sweep into striated ordering. In the Bloch sphere picture, $\phi = 0$ corresponds to the $+x$ axis, so the wavefunctions on (0,0) and (1,1) sublattices correspond to vectors pointing mostly up or mostly down with a small projection of each along the $+x$ axis. In the same Bloch sphere picture, quenching at $\phi_q = \pi/2$ or $-\pi/2$ corresponds to rotations around the $+y$ or $-y$ axes (Fig. \ref{fig_striated}a).\\

To resolve the local response of the system, we use high-order correlators which are extracted from single-shot site-resolved readout. In particular, we define an operator $\hat{\mathcal{O}}_i^{(d)}$ on the eight atoms surrounding site $i$. This operator projects the neighboring atoms into configurations in which all four nearest atoms are in $\ket{g}$ and exactly $d$ of the diagonal neighbors are in $\ket{r}$. Specifically, the operator $\hat{\mathcal{O}}_i^{(d)}$ decomposes into a projector $\hat{A}_i$ on the four nearest neighboring atoms and $\hat{B}_i^{(d)}$ on the four diagonal neighbors, according to $\hat{\mathcal{O}}_i^{(d)} = \hat{A}_i \hat{B}_i^{(d)}$. Defining $\bar{n}_i = |g\rangle_i\langle g|$ and $n_i = |r\rangle_i\langle r|$, the nearest neighbor projector is written as $\hat{A}_i = \prod_{\langle j, i \rangle } \bar{n}_j$, where $\langle . \rangle$ denotes nearest neighbors. The projector $\hat{B}_i^{(d)}$ sums over all configurations of the diagonal neighbors (indexed $k_1, k_2, k_3, k_4$) with $d$ excitations:
\begin{align}
\hat{B}_i^{(4)} &= n_{k_1}n_{k_2}n_{k_3}n_{k_4}  \\
\hat{B}_i^{(3)} &= \bar{n}_{k_1}n_{k_2}n_{k_3}n_{k_4} + n_{k_1}\bar{n}_{k_2}n_{k_3}n_{k_4} + \ldots \\
\hat{B}_i^{(2)} &= \bar{n}_{k_1}\bar{n}_{k_2}n_{k_3}n_{k_4} + \bar{n}_{k_1}{n}_{k_2}\bar{n}_{k_3}n_{k_4} + \ldots
\end{align}
These operators are used to construct the conditional Rydberg density  $$P^{(d)} = \frac{\sum_i \langle n_i \hat{\mathcal{O}}_i^{(d)}\rangle}{\sum_i \langle \hat{\mathcal{O}}_i^{(d)} \rangle}$$ which measures the probability of Rydberg excitation on site $i$ surrounded by neighboring-atom configurations for which $\hat{\mathcal{O}}_i^{(d)}=1$.

To quantify coherences, we measure these conditional probabilities on their corresponding resonances, after a fixed quench  with variable quench phase $\phi_q$. For a single particle driven by the Hamiltonian $H=\Omega (\cos \phi_q \sigma_x + \sin \phi_q \sigma_y)/2 + \Delta \sigma_z/2$ for time $\tau$, the resulting Heisenberg evolution is given by $\sigma_z' = U^\dagger \sigma_z U$, where $U = e^{-i H \tau}$. The resulting operator can be expressed as 
\begin{align}
    \sigma_z' &= \tilde{\Omega} \sin2\alpha (-\sigma_x \sin \phi_q + \sigma_y \cos \phi_q) \\
    &+ 2\tilde{\Delta}\tilde{\Omega} \sin^2\alpha (\sigma_x \cos\phi_q + \sigma_y \sin \phi_q) \\
    &+ (\cos^2 \alpha - (1 - 2\tilde{\Delta}^2)\sin^2 \alpha) \sigma_z
\end{align}
where $\tilde{\Delta} = \Delta / \sqrt{\Delta^2 + \Omega^2}$, $\tilde{\Omega} = \Omega / \sqrt{\Delta^2 + \Omega^2}$, and $\alpha = \frac{1}{2}\tau \sqrt{\Delta^2+\Omega^2}$.

We fit the conditional probabilites $P^{(0)}$ and $P^{(4)}$ as a function of $\phi_q$ (Fig.~\ref{fig_striated}d,e), taking $\Delta$ as the effective detuning from interaction-shifted resonance, and measuring $\langle \sigma_z' \rangle$ as a function of $\phi_q$ to extract the Bloch vector components $\langle \sigma_x \rangle, \langle \sigma_y \rangle, \langle \sigma_z \rangle$ on the two respective sublattices. For the (1,1) sublattice response, we model the evolution averaged over random detunings, due to $\sim 15\%$ fluctuations of the interaction shifts associated with thermal fluctuations in atomic positions, which broaden and weaken the spectroscopic response. For both sublattices we also include fluctuations in the calibrated pulse area ($\sim 10\%$ due to low power used). The extracted fit values are $\sigma_{x,y,z}^{(0,0)} = -0.82(6), 0.25(2), -0.32(4)$, and $\sigma_{x,y,z}^{(1,1)} = -0.46(4), 0.01(1), 0.91(5)$.\\


\noindent\textbf{Acknowledgements} We thank many members of the Harvard AMO community, particularly Elana Urbach, Samantha Dakoulas, and John Doyle for their efforts enabling safe and productive operation of our laboratories during 2020. We thank Hannes Bernien, Dirk Englund, Manuel Endres, Nate Gemelke, Donggyu Kim, Peter Stark, and Alexander Zibrov for discussions and experimental help. We acknowledge financial support from the Center for Ultracold Atoms, the National Science Foundation, the Vannevar Bush Faculty Fellowship, the U.S. Department of Energy, the Office of Naval Research, the Army Research Office MURI, and the DARPA ONISQ program. T.T.W. acknowledges support from Gordon College. H.L. acknowledges support from the National Defense Science and Engineering Graduate (NDSEG) fellowship. G.S. acknowledges support from a fellowship from the Max Planck/Harvard Research Center for Quantum Optics. D.B. acknowledges support from the NSF Graduate Research Fellowship Program (grant DGE1745303) and The Fannie and John Hertz Foundation. W.W.H. is supported by the Moore Foundation’s EPiQS Initiative Grant No. GBMF4306, the NUS Development Grant AY2019/2020, and the Stanford Institute of Theoretical Physics. S.C. acknowledges support from the Miller Institute for Basic Research in Science. R.S. and S.S. were supported by the U.S.~Department of Energy under Grant $\mbox{DE-SC0019030}$. The DMRG calculations were performed using the ITensor Library \cite{itensor}. The computations in this paper were run on the FASRC Cannon cluster supported by the FAS Division of Science Research Computing Group at Harvard University.\\

\noindent\textbf{Author contributions}
S.E., T.T.W., H.L., A.K., G.S, A. O. and D.B. 
contributed to the building  experimental setup, performed the measurements, and analyzed the data. Theoretical analysis was performed by R.S., H.P., W.W.H., and S.C. All work was supervised by S.S., M.G., V.V., and M.D.L. All authors discussed the results and contributed to the manuscript.\\

\noindent\textbf{Competing interests} M.G., V.V., and M.D.L. are co-founders and shareholders of QuEra Computing. A.O. is a shareholder of QuEra Computing.\\

\noindent\textbf{Correspondence and requests for materials} should be addressed to M.D.L.\\

\renewcommand{\figurename}{EXTENDED DATA FIG.}
\setcounter{figure}{0}

\begin{figure*}
\includegraphics{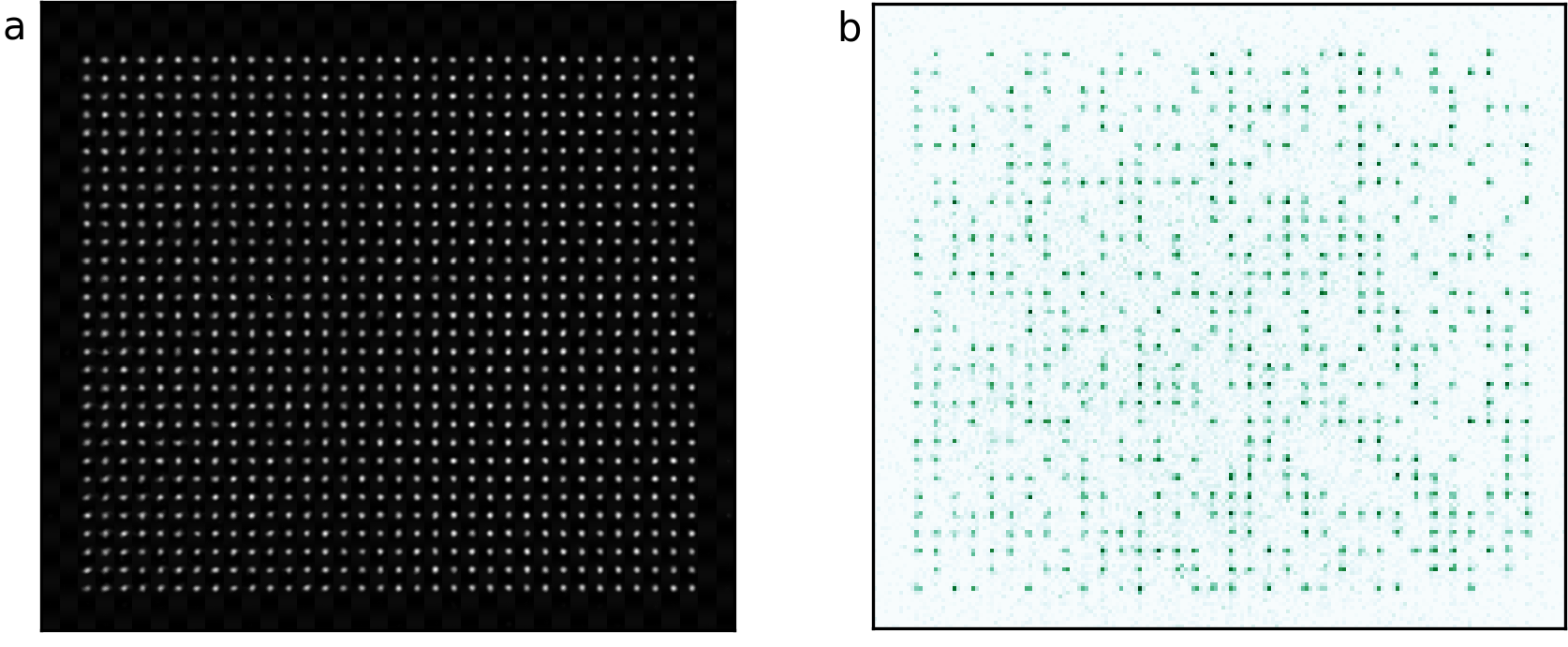}
\caption{\textbf{Large arrays of optical tweezers.} The experimental platform produces optical tweezer arrays with up to $\sim 1000$ tweezers and $\sim 50\%$ loading probability per tweezer after $100~$ms of MOT loading time. \textbf{a.} Camera image of an array of 34$\times$30 tweezers (1020 traps), including aberration correction. \textbf{b.} Sample image of random loading into this tweezer array, with 543 loaded atoms. Atoms are detected on an EMCCD camera with fluorescence imaging.}
\label{fig_large_tweezer_arrays}
\end{figure*}

\begin{figure*}
\includegraphics{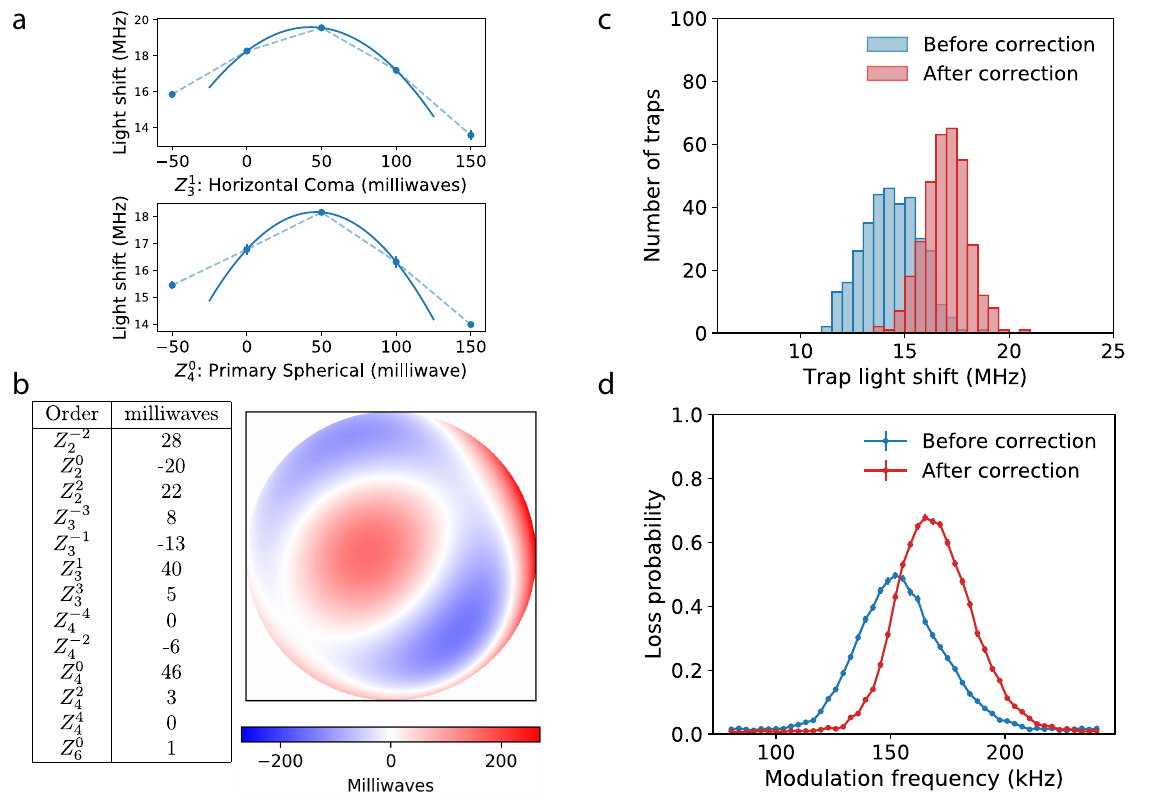}
\caption{\textbf{Correcting for aberrations in the SLM tweezer array.} The aberration correction procedure utilizes the orthogonality of Zernike polynomials and the fact that correcting aberrations increases tweezer light shifts on the atoms. To independently measure and correct each aberration type, Zernike polynomials are added with variable amplitude to the SLM phase hologram, with values optimized to maximize tweezer light shifts. \textbf{a.} Two common aberration types: horizontal coma (upper) and primary spherical (lower), for which $\sim 50$~milliwaves compensation on each reduces aberrations and results in higher-depth traps. \textbf{b.} Correcting for aberrations associated with the thirteen lowest order Zernike polynomials. The sum of all polynomials with their associated coefficients gives the total aberrated phase profile in the optical system, which is now corrected (total RMS aberration of $\sim 70$~milliwaves).
\textbf{c.} Trap depths across a $26\times13$ trap array before and after correction for aberrations. Aberration correction results in tighter focusing (higher trap light shift) and improved homogeneity. Trap depths are measured by probing the light shift of each trap on the $\ket{5S_{1/2}, F=2} \to \ket{5P_{3/2}, F'=2}$ transition. \textbf{d.} Aberration correction also results in higher and more homogeneous trap frequencies across the array. Trap frequencies are measured by modulating tweezer depths at variable frequencies, resulting in parametric heating and atom loss when the modulation frequency is twice the radial trap frequency. The measurement after correction for aberrations shows a narrower spectrum and higher trap frequencies (averaged over the whole array).}
\label{fig_tweezer_aberrations}
\end{figure*}

\begin{figure*}
\includegraphics{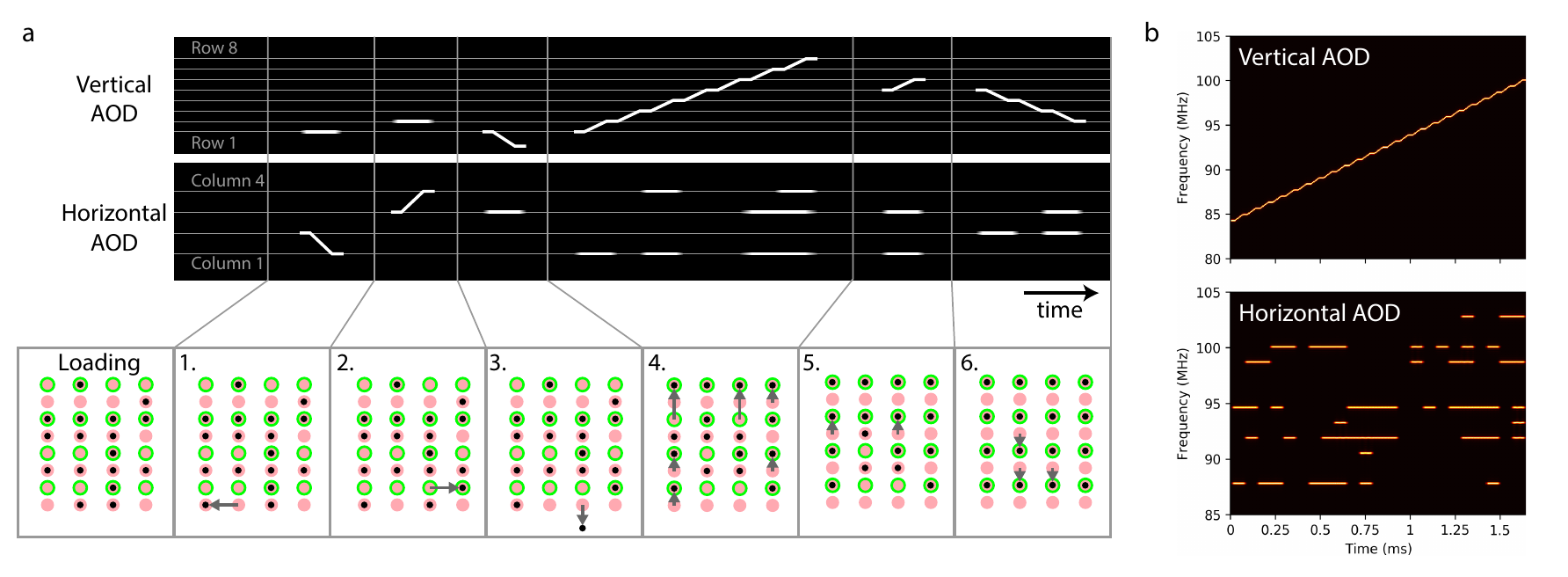}
\caption{\textbf{Rearrangement protocol.} \textbf{a.} Sample sequence of individual rearrangement steps. There are two pre-sorting moves (1, 2). Move (3) is the single ejection move. Moves (4-6) consist of parallel vertical sorting within each column, including both upward and downwards move. The upper panel illustrates the frequency spectrum of the waveform in the vertical and horizontal AODs during these moves, with the underlying grid corresponding to the calibrated frequencies which map to SLM array rows and columns. \textbf{b.} Spectrograms representing the horizontal and vertical AOD waveforms over the duration of a single vertical frequency scan during a realistic rearrangement procedure for a 26$\times$13 array. The heat-maps show frequency spectra of the AOD waveforms over small time intervals during the scan.}
\label{fig_rearrangement}
\end{figure*}

\begin{figure*}
\includegraphics[width=180mm]{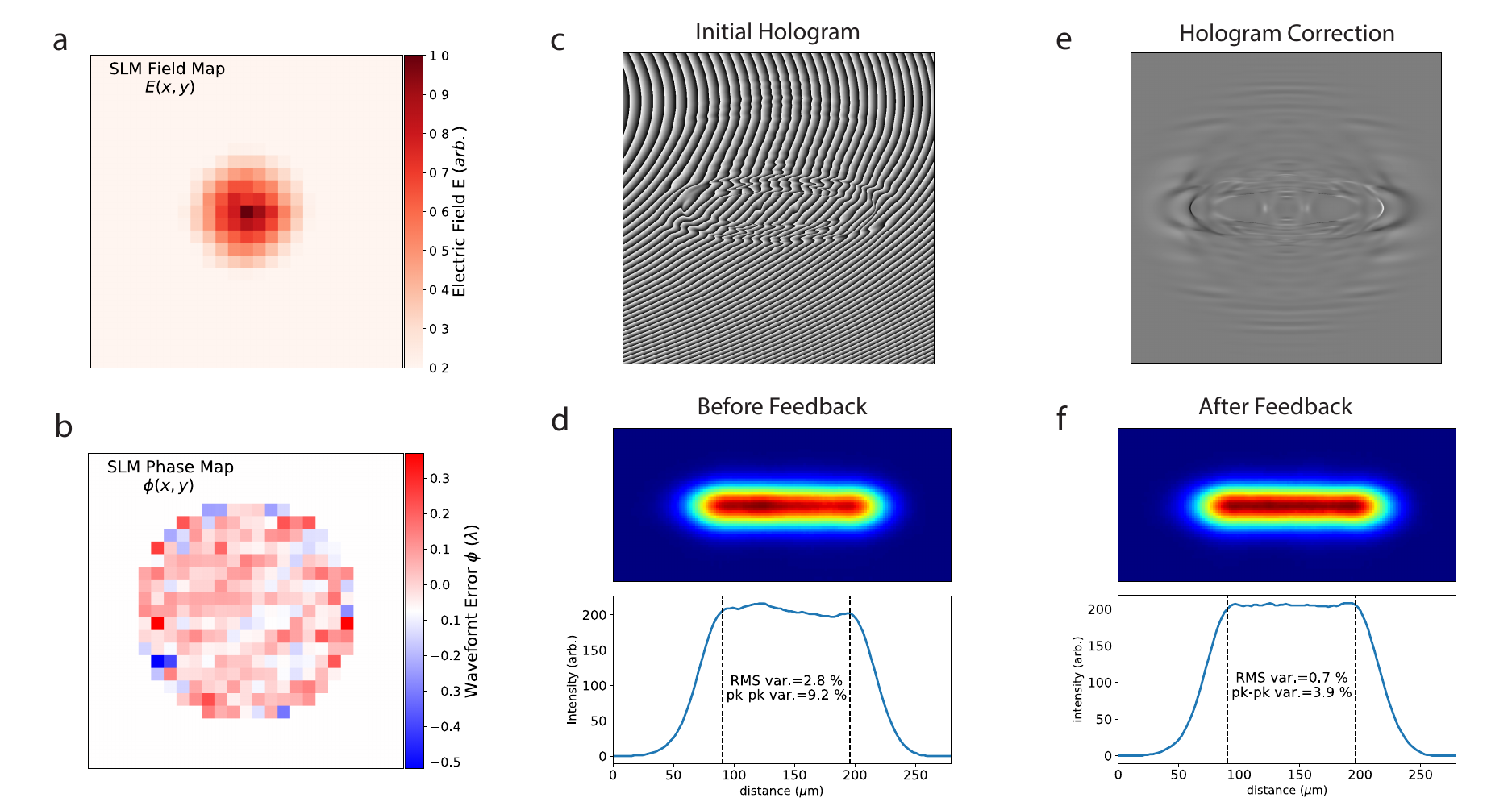}
\caption{\textbf{Generating homogeneous Rydberg beams} \textbf{a.} Measured Gaussian-beam illumination on the SLM for shaping the 420-nm Rydberg beam. A Gaussian fit to this data is used as an input for the hologram optimization algorithm. \textbf{b.} Corrected and measured wavefront error through our optical system, showing a reduction of aberrations to $\lambda/100$. \textbf{c.} Computer-generated hologram for creating the 420-nm top-hat beam. \textbf{d.} Measured light intensity of the 420-nm top-hat beam (top), and the cross section along where atoms will be positioned (bottom). Vertical lines denote the 105-$\mu$m region where the beam should be flat. \textbf{e.} Using the measured top-hat intensity, a phase correction is calculated for adding to the initial hologram. \textbf{f.} Resulting top-hat beam after feedback shows significantly improved homogeneity.}
\label{fig_tophat}
\end{figure*}

\begin{figure*}
\includegraphics{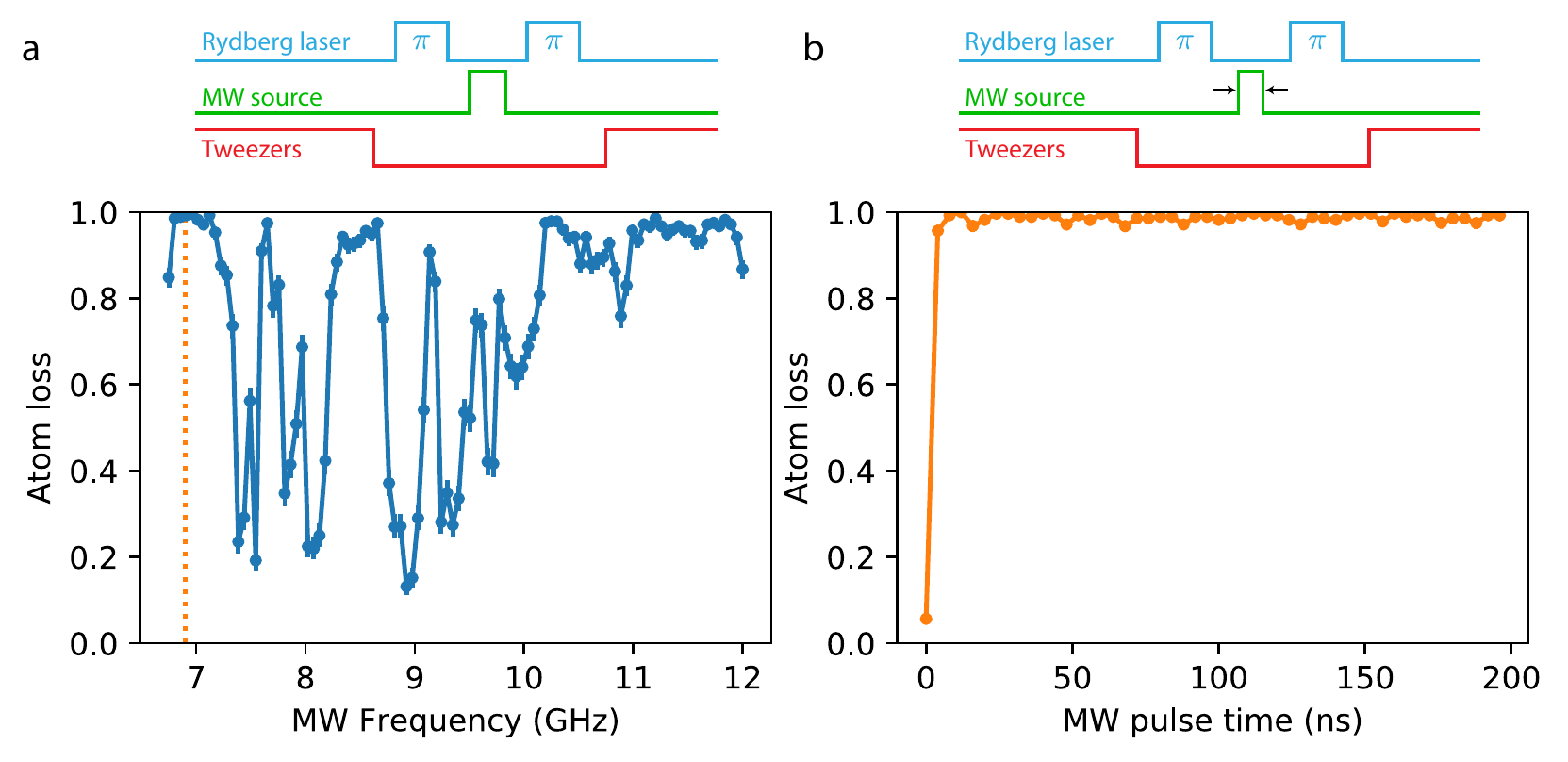}
\caption{\textbf{Characterizing microwave-enhanced Rydberg detection fidelity.} The effect of strong microwave (MW) pulses on Rydberg atoms is measured by preparing atoms in $\ket{g}$, exciting to $\ket{r}$ with a Rydberg $\pi$-pulse, and then applying the MW pulse before de-exciting residual Rydberg atoms with a final Rydberg $\pi$-pulse. (The entire sequence occurs while tweezers are briefly turned off.) \textbf{a.} Broad resonances are observed with varying microwave frequency, corresponding to transitions from $\ket{r} = \ket{70S}$ to other Rydberg states. Note that the transition to $\ket{69P}$ and $\ket{70P}$ are in the range of $10-12$~GHz, and over this entire range there is strong transfer out of $\ket{r}$. Other resonances might be due to multi-photon effects. \textbf{b.} With fixed $6.9$-GHz MW frequency and varying pulse time, there is a rapid transfer out of the Rydberg state on the timescale of several nanoseconds. Over short time-scales, there may be coherent oscillations which return population back to $\ket{r}$, so a $100$~ns pulse is used for enhancement of loss signal of $\ket{r}$ in the experiment.}
\label{fig_mw_ionization}
\end{figure*}

\begin{figure*}
\includegraphics[width=\textwidth]{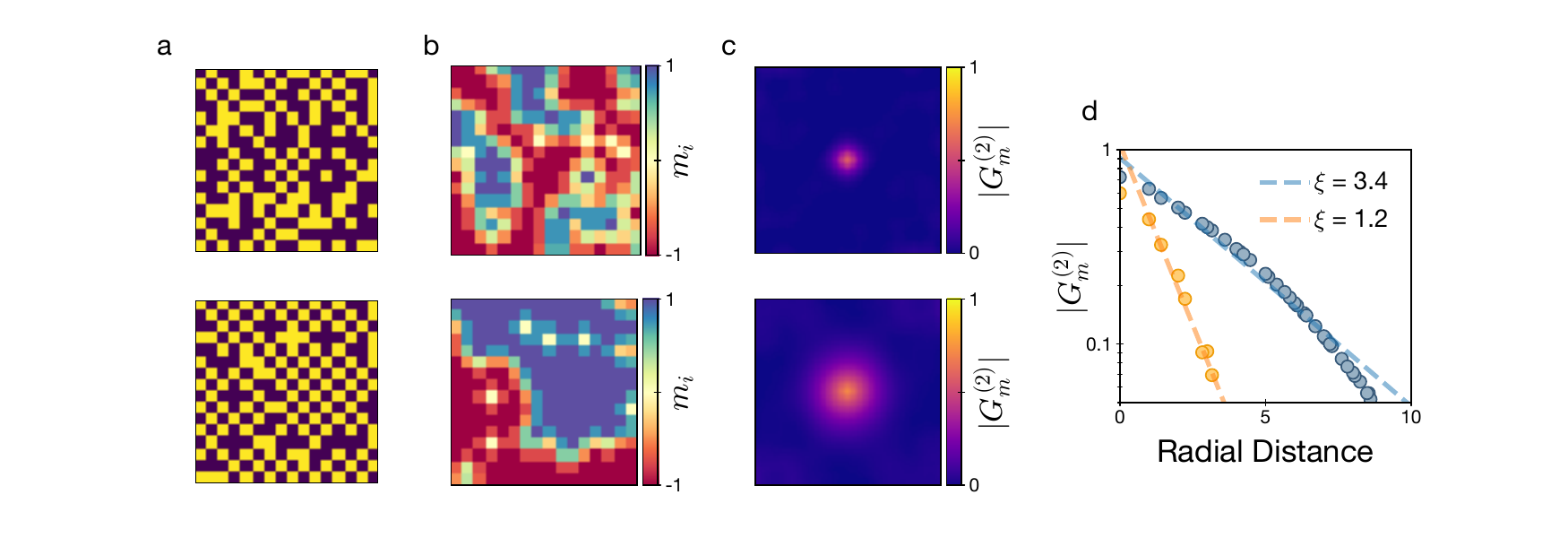}
\caption{\textbf{Coarse-grained local staggered magnetization.} \textbf{a.} Examples of Rydberg populations $n_i$ after a faster (top) and slower (bottom) linear sweep. \textbf{b.} Corresponding coarse-grained local staggered magnetizations $m_i$ clearly show larger extents of antiferromagnetically ordered domains (dark blue or dark red) for the slower sweep (bottom) compared to for the faster sweep (top), as expected from the Kibble-Zurek mechanism. \textbf{c.} Isotropic correlation functions $G^{(2)}_m$ for the corresponding coarse-grained local staggered magnetizations after a faster (top) or a slower (bottom). \textbf{d.} As a function of radial distance, correlations $G^{(2)}_m$ decay exponentially with a length scale corresponding to the correlation length $\xi$. The two decay curves correspond to faster (orange) and slower (blue) sweeps. 
}
\label{fig_local_stagg}
\end{figure*}

\begin{figure*}
\includegraphics[width=\textwidth]{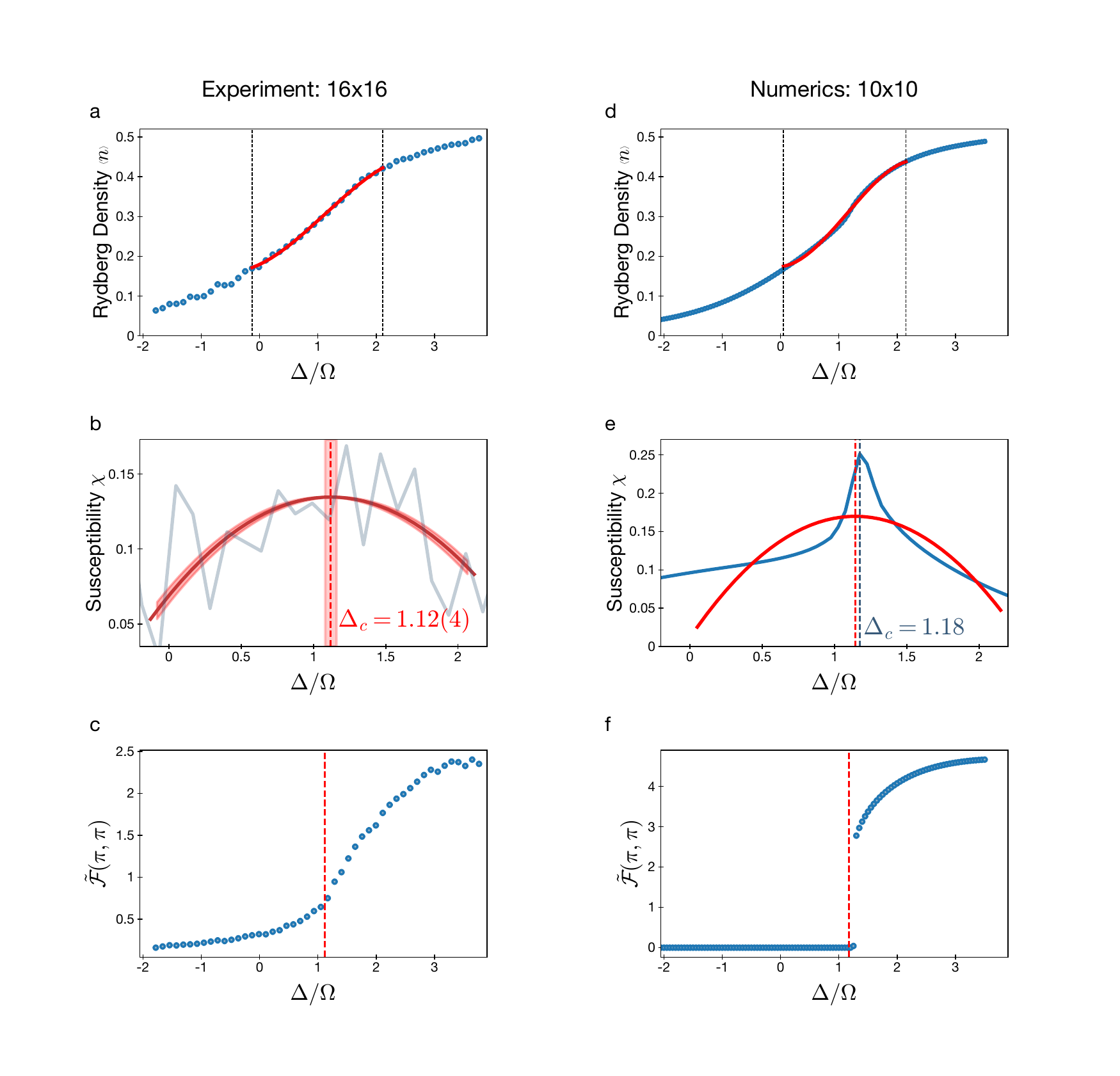}
\caption{\textbf{Extracting the quantum critical point.} \textbf{a.} The mean Rydberg excitation density $\langle n \rangle$ vs. detuning $\Delta/\Omega$ on a 16$\times$16 array. The data is fitted within a window (dashed lines) to a cubic polynomial (red curve) as a means of smoothening the data. \textbf{b.} The peak in the numerical derivative of the fitted data (red curve) corresponds to the critical point $\Delta_c/\Omega = 1.12(4)$ (red shaded regions show uncertainty ranges, obtained from varying fit windows). In contrast, the point-by-point slope of the data (gray) is too noisy to be useful. \textbf{c.} Order parameter $\tilde{\mathcal{F}}(\pi,\pi)$ for the checkerboard phase vs. $\Delta/\Omega$ measured on a 16$\times$16 array with the value of the critical point from \textbf{b.} superimposed (red line), showing the clear growth of the order parameter after the critical point. \textbf{d.} DMRG simulations of $\langle n \rangle$ vs. $\Delta/\Omega$ on a 10$\times$10 array. For comparison against the experimental fitting procedure, the data from numerics is also fitted to a cubic polynomial within the indicated window (dashed lines). \textbf{e.} The point-by-point slope of the numerical data (blue curve) has a peak at $\Delta_c/\Omega = 1.18$ (blue dashed line), in good agreement with the results (red dashed line) from both the numerical derivative of the cubic fit on the same data (red curve) and the result of the experiment. \textbf{f.} DMRG simulation of $\tilde{\mathcal{F}}(\pi,\pi)$ vs. $\Delta/\Omega$, with the exact quantum critical point from numerics shown (red line).}
\label{fig_sus}
\end{figure*}

\begin{figure*}
\includegraphics[width=\textwidth]{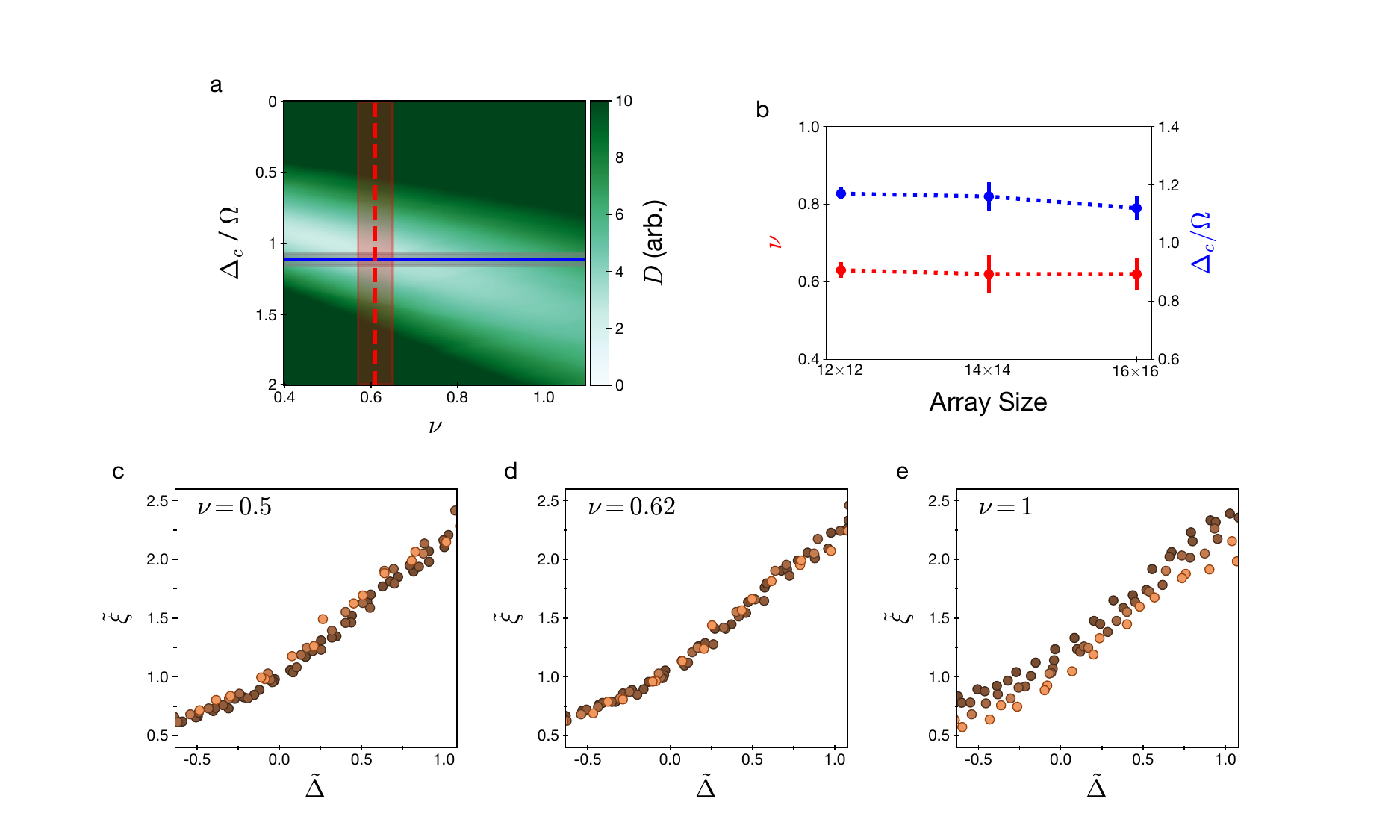}
\caption{\textbf{Optimization of data collapse.} \textbf{a.} Distance $D$ between rescaled correlation length $\tilde{\xi}$ vs. $\tilde{\Delta}$ curves depends on both the location of the quantum critical point location $\Delta_c/\Omega$ and on the correlation length critical exponent $\nu$. The independently determined $\Delta_c/\Omega$ (blue line, with uncertainty range in gray) and the experimentally extracted value of $\nu$ (dashed red line, with uncertainty range corresponding to the red shaded region) are marked on the plot. \textbf{b.} Our determination of $\nu$ (red) from data collapse around the independently determined $\Delta_c/\Omega$ (blue) is consistent across arrays of different sizes. \textbf{c-e.} Data collapse is clearly better at the experimentally determined value ($\nu=0.62$) as compared to the mean-field ($\nu=0.5$) or the (1+1)D ($\nu=1$) values. The horizontal extent of the data corresponds to the region of overlap of all rescaled data sets.}
\label{fig_collapse_distance}
\end{figure*}

\clearpage
\newpage





\bibliographystyle{apsrev4-1}
\bibliography{references3.bib}

\end{document}